\documentclass{elsarticle}

\usepackage[reviewcopy]{adndt}
\usepackage{longtable}

\usepackage{amsmath}

\usepackage{amssymb}

\biboptions{square,sort&compress}
\bibpunct[]{[}{]}{,}{n}{}{;}
\begin{document}

\begin{frontmatter}

\journal{Atomic Data and Nuclear Data Tables}


\title{Discovery of Zinc, Selenium, Bromine, and Neodymium Isotopes}

 \author{J. L. Gross}
 \author{J. Claes}
 \author{J. Kathawa}
 \author{M. Thoennessen\corref{cor1}}\ead{thoennessen@nscl.msu.edu}

 \cortext[cor1]{Corresponding author.}

 \address{National Superconducting Cyclotron Laboratory and \\ Department of Physics and Astronomy, Michigan State University, \\ East Lansing, MI 48824, USA}

\begin{abstract}
Currently, thirty-two zinc, thirty-two selenium, twenty-nine bromine and thirty-one neodymium isotopes have been observed and the discovery of these isotopes is discussed here. For each isotope a brief synopsis of the first refereed publication, including the production and identification method, is presented.
\end{abstract}

\end{frontmatter}





\newpage
\tableofcontents
\listofDtables

\vskip5pc

\section{Introduction}\label{s:intro}

The discovery of zinc, selenium, bromine, and neodymium isotopes is discussed as part of the series summarizing the discovery of isotopes, beginning with the cerium isotopes in 2009 \cite{2009Gin01}. Guidelines for assigning credit for discovery are (1) clear identification, either through decay-curves and relationships to other known isotopes, particle or $\gamma$-ray spectra, or unique mass and Z-identification, and (2) publication of the discovery in a refereed journal. The authors and year of the first publication, the laboratory where the isotopes were produced as well as the production and identification methods are discussed. When appropriate, references to conference proceedings, internal reports, and theses are included. When a discovery includes a half-life measurement the measured value is compared to the currently adopted value taken from the NUBASE evaluation \cite{2003Aud01} which is based on the ENSDF database \cite{2008ENS01}. In cases where the reported half-life differed significantly from the adopted half-life (up to approximately a factor of two), we searched the subsequent literature for indications that the measurement was erroneous. If that was not the case we credited the authors with the discovery in spite of the inaccurate half-life.

The first criterium excludes measurements of half-lives of a given element without mass identification. This affects mostly isotopes first observed in fission where decay curves of chemically separated elements were measured without the capability to determine their mass. Also the four-parameter measurements (see, for example, Ref. \cite{1970Joh01}) were, in general, not considered because the mass identification was only $\pm$1 mass unit.

The second criterium affects especially the isotopes studied within the Manhattan Project. Although an overview of the results was published in 1946 \cite{1946TPP01}, most of the papers were only published in the Plutonium Project Records of the Manhattan Project Technical Series, Vol. 9A, ''Radiochemistry and the Fission Products," in three books by Wiley in 1951 \cite{1951Cor01}. We considered this first unclassified publication to be equivalent to a refereed paper.

The initial literature search was performed using the databases ENSDF \cite{2008ENS01} and NSR \cite{2008NSR01} of the National Nuclear Data Center at Brookhaven National Laboratory. These databases are complete and reliable back to the early 1960's. For earlier references, several editions of the Table of Isotopes were used \cite{1940Liv01,1944Sea01,1948Sea01,1953Hol02,1958Str01,1967Led01}. A good reference for the discovery of the stable isotopes was the second edition of Aston's book ``Mass Spectra and Isotopes'' \cite{1942Ast01}.

\section{Discovery of $^{54-85}$Zn}

Thirty-two zinc isotopes from A = 54$-$85 have been discovered so far; these include 5 stable, 11 neutron-deficient and 16 neutron-rich isotopes. According to the HFB-14 model \cite{2007Gor01}, $^{93}$Zn should be the last odd-even particle stable neutron-rich nucleus while the even-even particle stable neutron-rich nuclei should continue through $^{106}$Zn. The proton dripline has been reached at $^{54}$Zn, but $^{53}$Zn could live long enough to be observed \cite{2004Tho01}. About 16 isotopes have yet to be discovered corresponding to 33\% of all possible neodymium isotopes.

\begin{figure}
	\centering
	\includegraphics[scale=.5]{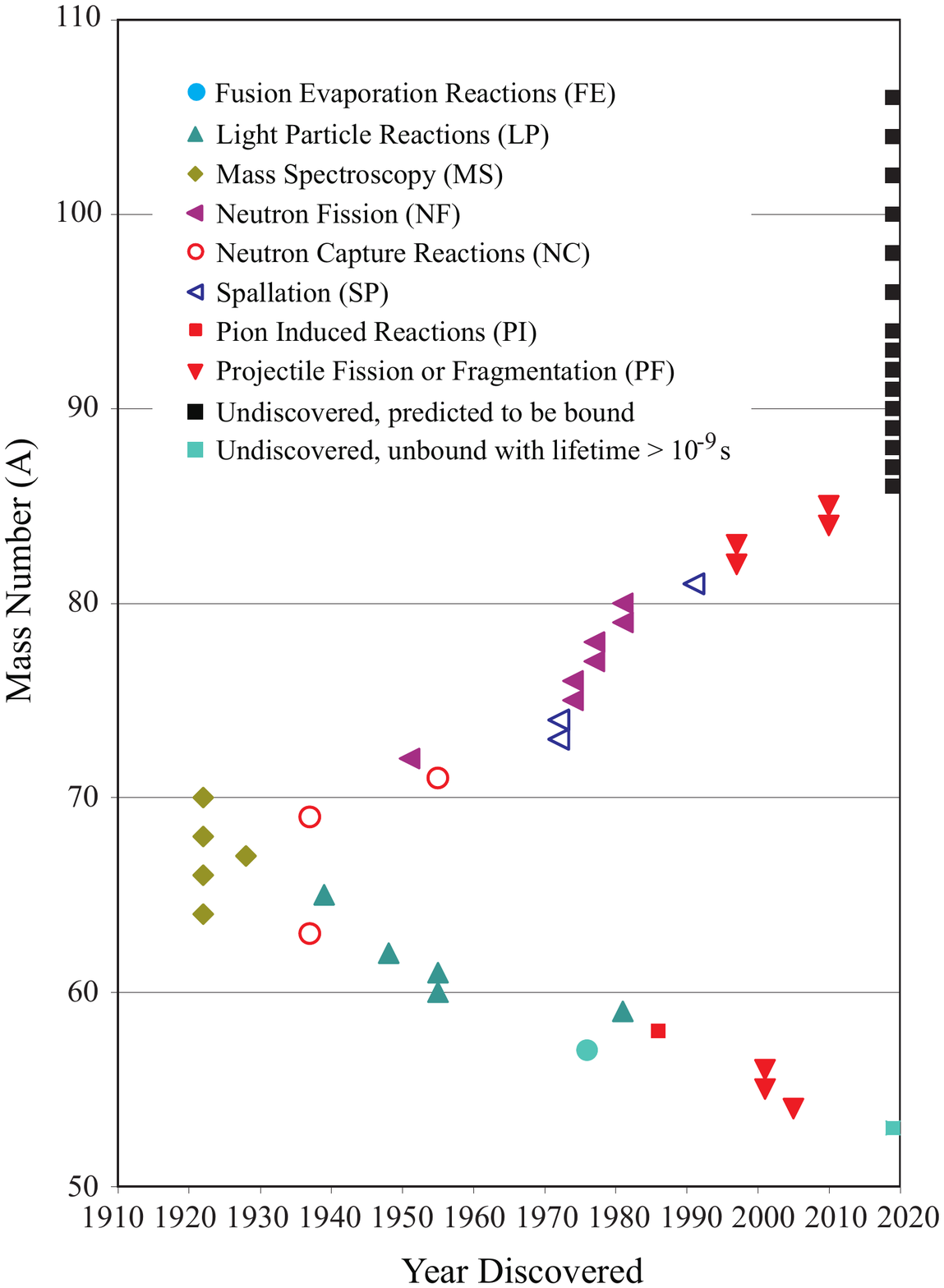}
	\caption{Zinc isotopes as a function of time when they were discovered. The different production methods are indicated. The solid black squares on the right hand side of the plot are isotopes predicted to be bound by the HFB-14 model. On the proton-rich side the light blue square correspond to an unbound isotope predicted to have lifetimes larger than $\sim 10^{-9}$~s.}
\label{f:year-zn}
\end{figure}

Figure \ref{f:year-zn} summarizes the year of first discovery for all zinc isotopes identified by the method of discovery. The range of isotopes predicted to exist is indicated on the right side of the figure. The radioactive zinc isotopes were produced using fusion evaporation reactions (FE), light-particle reactions (LP), neutron induced fission (NF), neutron-capture reactions (NC), spallation reactions (SP), pion-induced reactions (PI) and projectile fragmentation or fission (PF). The stable isotopes were identified using mass spectroscopy (MS). Light particles also include neutrons produced by accelerators. In the following, the discovery of each zinc isotope is discussed in detail and a summary is presented in Table 1.

\subsection{$^{54}$Zn}\vspace{0.0cm}

Blank et al. discovered $^{54}$Zn in 2005 and reported their results in ``First Observation of $^{54}$Zn and its Decay by Two-Proton Emission'' \cite{2005Bla01}. A $^{58}$Ni beam was accelerated to 74.5 MeV/nucleon by the GANIL cyclotrons and bombarded a natural nickel target. The fragments were separated by the ALPHA-LISE3 separator and identified with four silicon detectors. ``To be accepted all eight identification parameters of an event had to lie within 3 standard deviations of the predefined values. This procedure yields basically background free identification spectra. Thus, eight events have been attributed to $^{54}$Zn.'' The reported half-life of 3.2($^{+18}_{-8}$)~ms is the currently accepted half-life.

\subsection{$^{55,56}$Zn}\vspace{0.0cm}

$^{55}$Zn and $^{56}$Zn were first observed by Giovinazzo et al. in the 2001 paper, ``First Observation of $^{55,56}$Zn'' \cite{2001Gio01}. The GANIL cyclotrons accelerated a $^{58}$Ni beam to 74.5 MeV/nucleon, which then bombarded a natural nickel target. The isotopes were separated with the SISSI/LISE3 facility and identified in a three-element silicon implantation telescope. ``In order to accept a count as a valid event, each value of the parameters had to lie within a two-FWHM window of the determined central position. The results of this analysis is presented in [the figure]. We observe 17 counts of $^{56}$Zn and 14 counts of $^{55}$Zn.''

\subsection{$^{57}$Zn}\vspace{0.0cm}

Vieira et al. reported the observation of excited states of $^{57}$Zn in 1976 in the paper ``Extension of the T$_{z}=-$3/2 Beta-Delayed Proton Precursor Series to $^{57}$Zn'' \cite{1976Vie01}. The Berkeley 88-in. cyclotron accelerated $^{20}$Ne to 62 and 70 MeV which then bombarded natural calcium targets and $^{57}$Zn was produced in the fusion-evaporation reaction $^{40}$Ca($^{20}$Ne,3n). Beta-delayed protons were measured with a semiconducting counter telescope. ``These activities are, however, very compatible with the $\beta$-delayed proton emission expected from $^{57}$Zn whose predicted threshold via the $^{40}$Ca($^{20}$Ne,3n) reaction is $\approx$ 50.0 MeV$\ldots$'' The reported half-life of 40(10)~ms is currently the only measured value.

\subsection{$^{58}$Zn}\vspace{0.0cm}

$^{58}$Zn was first observed in 1986 by Seth et al. in ``Mass of Proton-Rich $^{58}$Zn by Pion Double Charge Exchange'' \cite{1986Set01}. A $\pi^+$ beam with a central energy of 292 MeV hit a $^{58}$Ni target to measure the ($\pi^+$,$\pi^-$) reaction at the Los Alamos Meson Physics Facility. $^{58}$Zn was identified by the detection of negative pions $\pi^-$ measured using a modified version of the EPICS spectrometer. ``Using the known atomic Q$_0=-17 565.6 \pm 2.2$ keV for the $^9$Be($\pi^+$,$\pi^-$)$^9$C(g.s.) reaction, we obtain the atomic Q$_0=-17 930 \pm 50$ keV for the $^{58}$Ni($\pi^+$,$\pi^-$)$^{58}$Zn(g.s.) reaction. This corresponds to an atomic mass excess of $-$42295 $\pm$ 50 keV for $^{58}$Zn.''

\subsection{$^{59}$Zn}\vspace{0.0cm}

Honkanen et al. identified $^{59}$Zn in the 1981 paper ``Beta Decay and Delayed Proton Emission of a New Nuclide $^{59}$Zn'' \cite{1981Hon01}. Enriched $^{58}$Ni was bombarded with 25 MeV tritons from the MC-20 cyclotron at the University of Jyv\"askyl\"a and $^{59}$Zn was produced in the reaction $^{58}$Ni($^3$He,2n). The recoil products were transported by a helium-jet system and deposited on a mylar tape where protons were measured with a Si(Au) surface barrier detector. ``The $\beta^+$ decay of a new nuclide $^{59}$Zn has been identified by $\beta$-delayed proton and $\gamma$-ray emission. Two $\gamma$-rays and 15 proton groups have been associated with the decay of $^{59}$Zn.'' The reported half-life of 210(20)~ms agrees with the currently accepted half-life of 182.0(18)~ms. Arai et al. submitted their independent identification of $^{59}$Zn less than three month later \cite{1981Ara01}.

\subsection{$^{60,61}$Zn}\vspace{0.0cm}

$^{60}$Zn and $^{61}$Zn were discovered by Lindner and Brinkman in their 1955 paper ``Zinc 60 and Zinc 61'' \cite{1955Lin01}. Nickel foils were irradiated with 52 MeV $\alpha$-particles from the Philips' cyclotron at the Instituut voor Kernphysisch Onderzoek in Amsterdam. The reaction products were chemically separated and decay curves were recorded using a $\gamma$-liquid counter and a $\beta$ counter. ``According to these experiments we conclude to the formation of the Zn isotopes $^{60}$Zn and $^{61}$Zn by the reactions $^{58}$Ni($\alpha$,n)$^{61}$Zn, $^{58}$Ni($\alpha$,2n)$^{60}$Zn and perhaps some $^{60}$Ni($\alpha$,3n)$^{61}$Zn, the latter two reactions being absent with 15 MeV He$^{2+}$-ions.'' The reported half-lives of 2.1(1)~min. ($^{60}$Zn) and 87(3)~s ($^{61}$Zn) agree with the presently adopted values of 2.38(5)~min. and 89.1(2)~s, respectively. Lindner and Brinkman mentioned earlier results from a conference abstract \cite{1955Cum01}. These results were published in a refereed journal only four years later \cite{1959Cum01}.

\subsection{$^{62}$Zn}\vspace{0.0cm}

``Products of High Energy Deuteron and Helium Ion Bombardments of Copper'' presented the first observation of $^{62}$Zn by Miller et al. in 1948 \cite{1948Mil01}. The bombardment of natural copper with 190 MeV deuterons from the Berkeley 184-inch frequency-modulated cyclotron was used to produce $^{62}$Zn. ``Removal of copper from the zinc fractions yielded in the copper fractions a pure 11-minute activity which was identified as Cu$^{62}$; the 9.5-hour activity was therefore assigned to Zn$^{62}$.'' This half-life is consistent with the presently accepted value of 9.186(13)~h.

\subsection{$^{63}$Zn}\vspace{0.0cm}

Bothe and Gentner observed $^{63}$Zn in 1937 in ``Weitere Atomumwandlungen durch $\gamma$-Strahlen'' \cite{1937Bot03}. Lithium-$\gamma$-rays bombarded zinc targets and $^{63}$Zn was identified by assuming photo-nuclear reactions. ``Zink: T = 38 min. F\"ur Zn$^{65}$ hat Heyn T = 60 min gemessen. Wenn man nicht annehmen will, da\ss\ diese Bestimmung ziemlich ungenau ist, mu\ss\ unsere Halbwertzeit dem neuen Isotop Zn$^{63}$ zugeordnet werden.'' [Zinc: T = 38 min. Heyn measured a 60 min. half-life for Zn$^{65}$. If one does not want to assume, that this determination is rather inaccurate, this half-life has to be assigned to Zn$^{63}$.] This half-life agrees with the currently accepted value of 38.47(5)~min. Previously there had been an incorrect report that $^{63}$Zn was stable \cite{1935Ste01}.

\subsection{$^{64}$Zn}\vspace{0.0cm}

$^{64}$Zn was first observed by Dempster in 1922. He reported his result in ``Positive-ray Analysis of Potassium, Calcium and Zinc'' \cite{1922Dem01}. Zinc isotopes were identified with the positive-ray apparatus using a zinc anode. ``It was at first assumed that the atomic weights of the components were 63, 65, 67 and 69, since, with the intensities first observed, a mean atomic weight was obtained which agreed with the chemical atomic weight. This reasoning was invalidated by the different intensity ratios obtained with the improved shielding, and a direct comparison with the calcium component at 40 showed the atomic weights to be near the integers 64, 66, 68, and 70.'' The incorrect assignment had been published a year earlier \cite{1921Dem01}.

\subsection{$^{65}$Zn}\vspace{0.0cm}

Livingood and Seaborg identified $^{65}$Zn correctly in ``Radioactive Isotopes of Zinc'' in 1939 \cite{1939Liv03}. Zinc targets were irradiated by 8 MeV deuterons at the Berkeley cyclotron and the activities were measured with an air-filled quartz-fiber electroscope. ``It is shown that Zn$^{65}$, produced through deuteron bombardment of zinc, has a half-life of 250 days, agreeing with the period found by others after proton and deuteron bombardment of copper.'' The reported half-life of 250(5)~d agrees with the currently accepted value of 244.26(26)~d. A half-life of 245~d \cite{1938Per01} was reported previously without a mass assignment and half-lives of $\sim$1~h \cite{1936Hey01} and 210(3)~min \cite{1938Sag01} were incorrectly assigned to $^{65}$Zn. Also, $^{65}$Zn had been incorrectly observed to be stable \cite{1928Ast01,1935Ste01}.

\subsection{$^{66}$Zn}\vspace{0.0cm}

$^{66}$Zn was first observed by Dempster in 1922. He reported his result in ``Positive-ray Analysis of Potassium, Calcium and Zinc'' \cite{1922Dem01}. Zinc isotopes were identified with the positive-ray apparatus using a zinc anode. ``It was at first assumed that the atomic weights of the components were 63, 65, 67 and 69, since, with the intensities first observed, a mean atomic weight was obtained which agreed with the chemical atomic weight. This reasoning was invalidated by the different intensity ratios obtained with the improved shielding, and a direct comparison with the calcium component at 40 showed the atomic weights to be near the integers 64, 66, 68, and 70.'' The incorrect assignment had been published a year earlier \cite{1921Dem01}.

\subsection{$^{67}$Zn}\vspace{0.0cm}

Aston reported the observation of $^{67}$Zn in his 1928 paper, ``The Constitution of Zinc'' \cite{1928Ast01}. Zinc methyl discharge was used to identify the stable isotopes of zinc. ``The mass-spectra indicate that zinc consists of seven isotopes 64(a), 65(e), 66(b), 67(d), 68(c), 69(g), 70(f). The letters in brackets indicate the order of intensity.'' It should be mentioned that the observations of $^{65}$Zn and $^{69}$Zn were incorrect. Dempster suggested the existence of $^{67}$Zn six years earlier \cite{1922Dem01} but was not able to resolve the peak between $^{66}$Zn and $^{68}$Zn.

\subsection{$^{68}$Zn}\vspace{0.0cm}

$^{68}$Zn was first observed by Dempster in 1922. He reported his result in ``Positive-ray Analysis of Potassium, Calcium and Zinc'' \cite{1922Dem01}. Zinc isotopes were identified with the positive-ray apparatus using a zinc anode. ``It was at first assumed that the atomic weights of the components were 63, 65, 67 and 69, since, with the intensities first observed, a mean atomic weight was obtained which agreed with the chemical atomic weight. This reasoning was invalidated by the different intensity ratios obtained with the improved shielding, and a direct comparison with the calcium component at 40 showed the atomic weights to be near the integers 64, 66, 68, and 70.'' The incorrect assignment had been published a year earlier \cite{1921Dem01}.

\subsection{$^{69}$Zn}\vspace{0.0cm}

Heyn reported the observation of $^{69}$Zn in ``The Radioactivity of Nickel, Copper and Zinc'' in 1937 \cite{1937Hey02}. Zinc was irradiated by fast and slow neutrons and $\beta$-rays were magnetically analyzed following chemical separation. ``In zinc bombarded with slow neutrons we observed an activity with a period of 60 minutes. As the charge of the particles emitted proved to be negative, and as the activity is very strong and cannot be obtained by fast neutrons or gamma-rays, the carrier of this activity is most probably Zn$^{69}$.'' The observed half-life of 60~min. agrees with the currently accepted value of 56.4(9)~min. Heyn had previously assigned this activity incorrectly to $^{65}$Zn \cite{1936Hey01}. A 1.0(3)~h half-life was measured a year earlier without a definite mass assignment \cite{1936Liv01}. $^{69}$Zn had also been incorrectly observed to be stable \cite{1928Ast01}.

\subsection{$^{70}$Zn}\vspace{0.0cm}

$^{70}$Zn was first observed by Dempster in 1922. He reported his result in ``Positive-ray Analysis of Potassium, Calcium and Zinc'' \cite{1922Dem01}. Zinc isotopes were identified with the positive-ray apparatus using a zinc anode. ``It was at first assumed that the atomic weights of the components were 63, 65, 67 and 69, since, with the intensities first observed, a mean atomic weight was obtained which agreed with the chemical atomic weight. This reasoning was invalidated by the different intensity ratios obtained with the improved shielding, and a direct comparison with the calcium component at 40 showed the atomic weights to be near the integers 64, 66, 68, and 70.'' The incorrect assignment had been published a year earlier \cite{1921Dem01}.

\subsection{$^{71}$Zn}\vspace{0.0cm}

In 1954, LeBlanc et al. detected $^{71}$Zn as reported in ``Radioactivities of Zn$^{69}$ and Zn$^{71}$'' \cite{1955LeB01}. Normal zinc and enriched $^{70}$Zn were irradiated by neutrons at Argonne National Laboratory and $^{71}$Zn was identified with a ten-channel scintillation coincidence spectrometer. ``In all, four activities were found in the enriched Zn$^{70}$ samples. They had half-lives of 2.2~min, 1~hr, 3~hr, and 14~hr. Of these, only the 1-hr and 14-hr activities were detected in the normal Zn sources and are thus identified as the previously reported activities of Zn$^{69}$. The 2.2-min and 3-hr activities must then be due to Zn$^{71}$.'' These half-lives agree with the currently accepted values of 2.45(10)~m for the ground state and 3.96(5)~h for a long-lived isomer.

\subsection{$^{72}$Zn}\vspace{0.0cm}

The first identification of $^{72}$Zn was reported by Siegel and Glendenin in 1951 in ``Zinc and gallium activities in uranium fission'' \cite{1951Sie01} as part of the Manhattan Project as summarized in 1946 \cite{1946TPP01}. Dissolved uranium metal was irradiated in the Clinton Pile and decay and absorption curves were measured following chemical separation. ``From the decay rate of the Zn$^{72} - $Ga$^{72}$ pair in transient equilibrium, the half-life of Zn$^{72}$ was found to be 49.0 $\pm$ 1 hr over a period of about eight half-lives.'' This half-life agrees with the presently adopted value of 46.5(1)~hr.

\subsection{$^{73,74}$Zn}\vspace{0.0cm}

Erdal et al. published the observation of $^{73}$Zn and $^{74}$Zn in ``New Isotopes $^{73}$Zn and $^{74}$Zn'' \cite{1972Erd01} in 1972. At CERN, 600 MeV protons were used to bombard molten germanium and the isotopes were identified at the ISOLDE facility. Activities were collected on a moving-tape system and moved to a detector or to aluminum strips for off-line measurements. ``From a least-squares analysis, assuming two decay components, the half-life of $^{73}$Zn is determined to be 23.5 $\pm$ 1.0~sec.$\ldots$ The half-life of $^{74}$Zn was found to be 98 $\pm$ 2 sec, from a multiscaling analysis of the gross $\beta$-decay, in agreement with the value extracted from the $\gamma$-multi-analysis measurements$\ldots$'' The half-life of $^{73}$Zn corresponds to the accepted value and the half-life of $^{74}$Zn is included in the determination of the currently adopted value.

\subsection{$^{75,76}$Zn}\vspace{0.0cm}

$^{75}$Zn and $^{76}$Zn were observed by Grapengiesser et al. in the 1974 paper ``Survey of Short-lived Fission Products Obtained Using the Isotope-Separator-On-Line Facility at Studsvik'' in 1974 \cite{1974Gra01}. The zinc isotopes were produced by neutron induced fission and identified at the OSIRIS isotope-separator online facility at the Studsvik Neutron Research Laboratory in Nyk\"oping, Sweden. In the first long table, the half-life of $^{75}$Zn is quoted as 10.2(3)~s, which agrees with the currently accepted value of 10.2(2)~s. The half-life of $^{76}$Zn is quoted as 5.7(3)~s, which is the currently accepted half-life.

\subsection{$^{77,78}$Zn}\vspace{0.0cm}

Aleklett et al. reported the observation of $^{77}$Zn and  $^{78}$Zn in ``Total $\beta$-Decay Energies and Masses of Short-Lived Isotopes of Zinc, Gallium, Germanium and Arsenic'' in 1977 \cite{1977Ale01}. Fission products from the R2-0 reactor at Studsvik were detected by the OSIRIS separator facility. ``The $\gamma$-lines depopulating lower levels have also been used as gates but the coincident $\beta$-spectra have the same end-point energies as those feeding the levels around 2 MeV thus proving the large $\beta$-feeding to this region... The resulting Q$_\beta$ value for $^{77}$Zn is Q$_\beta$ = 6.91 $\pm$ 0.22 MeV... The half-life 1.6~s for $^{78}$Zn has been determined by means of $\gamma$-counting.'' This half-life agrees with the currently accepted value of 1.47(15)~s. The half-life of 1.4~s quoted for $^{77}$Zn had been mentioned in earlier publications \cite{1974Gra01,1975Ale01}, however, these papers were only referring to unpublished work. In addition, this half-life was incorrect, falling between the value for the ground state of 2.08(5)~s and an isomeric excited state of 1.05(10)~s.

\subsection{$^{79,80}$Zn}\vspace{0.0cm}

``Chemical Separation Combined with an ISOL-System'' by Rudstam et al. in 1981 \cite{1981Rud01} was the first report of $^{79}$Zn and $^{80}$Zn. $^{79,80}$Zn were produced by neutron induced fission and identified at the OSIRIS isotope-separator online facility at the Studsvik Neutron Research Laboratory in Nyk\"oping, Sweden. Gamma-ray spectra were measured with a Ge(Li)-spectrometer following mass and chemical separation.  ``Activities were found at all mass numbers from 76 (lowest one checked) to 80.'' No half-lives were reported but $\gamma$-rays in the daughter nuclei $^{79}$Ga and $^{80}$Ga were correctly identified. In two previous papers Rudstam et al. reported a half-life of 2.63(9)~s but they were not able to uniquely identify the element and assigned it to $^{79}$(Zn,Ga) \cite{1976Rud01,1977Rud01}.

\subsection{$^{81}$Zn}\vspace{0.0cm}

The discovery of $^{81}$Zn was reported by Kratz et al. in ``Neutron-rich isotopes around the \textit{r}-process `waiting-point' nuclei $^{79}_{29}$Cu$_{50}$ and $^{80}_{30}$Zn$_{50}$'' in 1991 \cite{1991Kra01}. A $^{238}$UC-graphite target was irradiated with 600 MeV protons from the CERN synchro-cyclotron and the fragments were separated and identified with the ISOLDE on-line mass separator. ``During the experiment, three further new isotopes could be identified, i.e. $^{77}$Cu, $^{81}$Zn, and $^{84}$Ga, the latter two lying even `beyond' the r-process path...'' The reported half-life of 290(50)~ms is included in the currently accepted value of 0.32(5)~s.

\subsection{$^{82,83}$Zn}\vspace{0.0cm}

Bernas et al. observed $^{82}$Zn and $^{83}$Zn for the first time in 1997 as reported in their paper ``Discovery and cross-section measurement of 58 new fission products in projectile-fission of 750$\cdot$A MeV $^{238}$U'' \cite{1997Ber01}. Uranium ions were accelerated to 750 A$\cdot$MeV by the GSI UNILAC/SIS accelerator facility and bombarded a beryllium target. The isotopes produced in the projectile-fission reaction were separated using the fragment separator FRS and the nuclear charge Z for each was determined by the energy loss measurement in an ionization chamber. ``The mass identification was carried out by measuring the time of flight (TOF) and the magnetic rigidity B$\rho$ with an accuracy of 10$^{-4}$.'' 125 and 10 counts of $^{82}$Zn and $^{83}$Zn were observed, respectively.

\subsection{$^{84,85}$Zn}\vspace{0.0cm}

The discovery of $^{84}$Zn and $^{85}$Zn was reported in the 2010 article ``Identification of 45 New Neutron-Rich Isotopes Produced by In-Flight Fission of a $^{238}$U Beam at 345 MeV/nucleon,'' by Ohnishi et al. \cite{2010Ohn01}. The experiment was performed at the RI Beam Factory at RIKEN, where the new isotopes were created by in-flight fission of a 345 MeV/nucleon $^{238}$U beam on a beryllium target. $^{84}$Zn and $^{85}$Zn were separated and identified with the BigRIPS superconducting in-flight separator. The list of new isotopes discovered in this study are summarized in a table. Twenty-two individual counts for $^{84}$Zn and one count for $^{85}$Zr were recorded.

\section{Discovery of $^{64-95}$Se}

Thirty-two selenium isotopes from A = $64-95$ have been discovered so far; these include 6 stable, 11 proton-rich and 15 neutron-rich isotopes.  According to the HFB-14 model \cite{2007Gor01}, $^{111}$Se should be the last odd-even particle stable neutron-rich nucleus while the even-even particle stable neutron-rich nuclei should continue through $^{118}$Se. At the proton dripline $^{62}$Se and $^{63}$Se could still be particle stable and $^{61}$Se could live long enough to be observed \cite{2004Tho01}. About 23 isotopes have yet to be discovered corresponding to 42\% of all possible selenium isotopes.

\begin{figure}
	\centering
	\includegraphics[scale=.5]{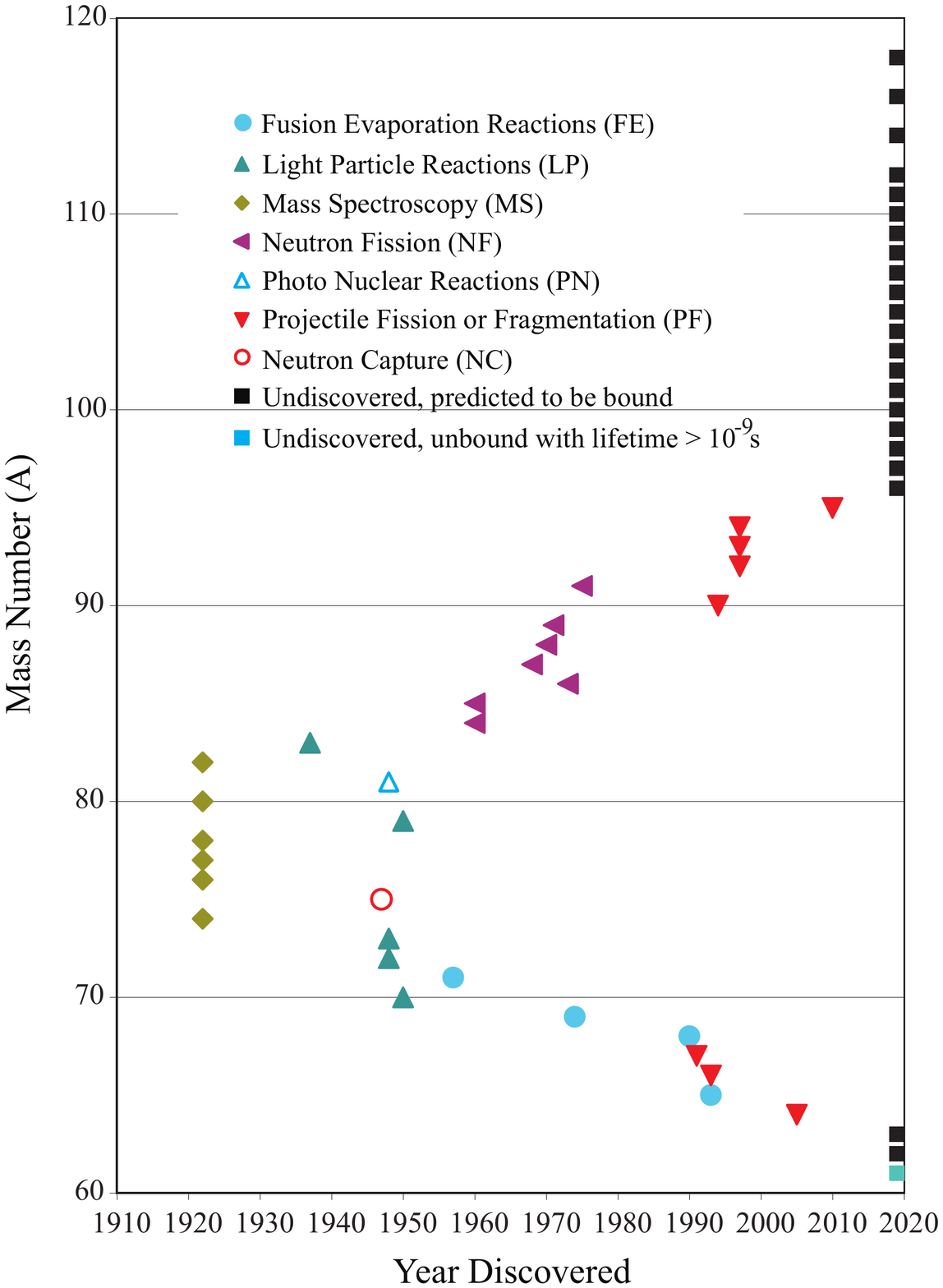}
	\caption{Selenium isotopes as a function of time when they were discovered. The different production methods are indicated. The solid black squares on the right hand side of the plot are isotopes predicted to be bound by the HFB-14 model. On the proton-rich side the light blue square correspond to $^{61}$Se which is predicted to have a lifetime larger than $\sim 10^{-9}$~s.}
\label{f:year-se}
\end{figure}

Figure \ref{f:year-se} summarizes the year of first discovery for all selenium isotopes identified by the method of discovery. The range of isotopes predicted to exist is indicated on the right side of the figure. The radioactive selenium isotopes were produced using heavy-ion fusion-evaporation reactions (FE), light-particle reactions (LP), neutron capture (NC), neutron-induced fission (NF), photo-nuclear reactions (PN) and projectile fragmentation or fission (PF). The stable isotopes were identified using mass spectroscopy (MS). Heavy ions are all nuclei with an atomic mass larger than A=4 \cite{1977Gru01}. Light particles also include neutrons produced by accelerators. In the following, the discovery of each selenium isotope is discussed in detail and a summary is presented in Table 1.

\subsection{$^{64}$Se}\vspace{0.0cm}
In the 2005 paper ``First observation of $^{60}$Ge and $^{64}$Se'' Stolz et al. identified the isotope $^{64}$Se for the first time \cite{2005Sto01}. $^{64}$Se was produced in the projectile fragmentation reaction of a 140 MeV/nucleon $^{78}$Kr beam on a beryllium target at the Coupled Cyclotron Facility of the National Superconducting Cyclotron Laboratory at Michigan State University. The projectile fragments were identified with the A1900 fragment separator. ``A total of four events of $^{64}$Se were observed during 32 hours of beam on target with an average primary beam current of 13.5~pnA.''

\subsection{$^{65}$Se}\vspace{0.0cm}
Batchelder et al. first observed $^{65}$Se as reported in the 1993 paper ``Beta-delayed proton decay of $^{65}$Se'' \cite{1993Bat01}. The Berkeley 88-Inch Cyclotron was used to accelerate a 175 MeV $^{28}$Si beam which bombarded a natural calcium target. $^{65}$Se was produced in the fusion-evaporation reaction $^{40}$Ca($^{28}$Si,3n) and the recoil products were deposited on a moving tape collector with a helium-jet setup. Beta-delayed protons were detected with a Si-Si detector telescope. ``A single proton group at 3.55$\pm$0.03 MeV has been observed... Combining this measurement with a Coulomb displacement energy calculation yields a mass excess for $^{65}$Se of $-$33.41$\pm$0.26 MeV.''

\subsection{$^{66}$Se}\vspace{0.0cm}
In 1993, Winger et al. reported the discovery of $^{66}$Se in ``Half-life measurements for $^{61}$Ga, $^{63}$Ge, and $^{65}$As and their importance in the rp process'' \cite{1993Win01}.  A 75 MeV/nucleon $^{78}$Kr beam from the Michigan State Cyclotron bombarded an enriched $^{58}$Ni. $^{66}$Se was separated and identified with the A1200 fragment separator. ``By using the triple gating method, we were able to observe 16 implantation events positively identified as $^{66}$Se and sufficient to demonstrate first observation of this proton-rich nucleus.''

\subsection{$^{67}$Se}\vspace{0.0cm}
Mohar et al. first observed $^{67}$Se in the 1991 paper ``Identification of New Nuclei near the Proton-Dripline for 31$\leq$Z$\leq$38'' \cite{1991Moh01}. A 65~A$\cdot$MeV $^{78}$Kr beam produced by the Michigan State K1200 cyclotron reacted with a $^{58}$Ni target. $^{67}$Se was identified by measuring the rigidity, $\Delta$\textit{E}, \textit{E}$_{total}$, and velocity in the A1200 fragment separator. ``Several new isotopes at or near the proton-drip line are indicated in the mass spectra: $^{61}$Ga, $^{62}$Ge, $^{63}$Ge, $^{65}$As, $^{69}$Br, and $^{75}$Sr.'' The discovery of $^{67}$Se was not explicitly mentioned but it is clearly identified in the selenium mass spectrum. The authors assumed it to be known, although it was only reported in a conference abstract \cite{1989Lan01}.

\subsection{$^{68}$Se}\vspace{0.0cm}
In 1990, Lister et al. reported the observation of $^{68}$Se in the paper ``Shape Changes in N=Z nuclei from Germanium to Zirconium'' \cite{1990Lis01}.  A 175 MeV $^{58}$Ni beam from the Daresbury NSF tandem accelerator bombarded a carbon target and $^{68}$Se was produced in the fusion-evaporation reaction $^{12}$C($^{58}$Ni,2n). Gamma-rays measured with ten shielded germanium were recorded in coincidence with recoil products measured in an ionization chamber. ``The final spectrum from our experiment on $^{68}$Se is shown in [the figure]. Although the statistics are poor, several features are clear. The strongest transition is at 853.9$\pm$0.3 keV and is a candidate for the 2$^+ \rightarrow$ 0$^+$ decay.'' Previous measurements of the $^{68}$Se half-life of 3.2(2)~h \cite{1972Bil01} and 1.6(4)~min. \cite{1976LaB01} were evidently incorrect.

\subsection{$^{69}$Se}\vspace{0.0cm}
Nolte et al. identified $^{69}$Se in the 1974 publication ``Investigation of Neutron Deficient Nuclei in the Region 28$<$N, Z$<$50 with the Help of Heavy Ion Compound Reactions'' \cite{1974Nol01}. The Munich MP tandem accelerator was used to bombard calcium targets with $^{32}$S beams of 90 and 100 MeV. $^{69}$Se was produced in the fusion-evaporation reaction $^{40}$Ca($^{32}$S,2pn) and identified with $\gamma$-ray and activation measurements. ``The half-life of the new isotope $^{69}$Se has been found to be 27$\pm$3 sec.'' This half-life agrees with the currently adopted value of 27.4(2)~s. A previous measurement of 1.8 and 14~min. isomeric states \cite{1973Pre01} was evidently incorrect.

\subsection{$^{70}$Se}\vspace{0.0cm}
In the 1950 paper ``Spallation Products of Arsenic with 190~MeV Deuterons'' Hopkins identified the isotope $^{70}$Se \cite{1950Hop01}. A pure $^{75}$As target was bombarded with 190~MeV deuterons from the Berkeley 184-inch cyclotron and chemically separated and subjected to spectrographic analysis. ``Table 1 contains two changes in isotope assignment differing from those previously reported. The 44-min. selenium and 52-min. arsenic daughter are placed at mass 70 since careful separations revealed no active germanium daughter.''  The measured half-life of 44~min. is in good agreement with the presently accepted value of 41.1(3)~min. In a previous paper the activity was incorrectly assigned to $^{71}$Se \cite{1948Hop01}.

\subsection{$^{71}$Se}\vspace{0.0cm}
Beydon et al reported $^{71}$Se in the 1957 publication ``Mise en \'evidence d'un isotope noveau de s\'elenium d\'eficient en neutrons'' \cite{1957Bey01}. Nickel and Copper targets were bombarded with a $^{14}$N beam from the Saclay cyclotron. Following chemical separation $^{71}$Se was identified by measuring the activity with a Geiger-M\"uller counter and NaI(Tl) detector. ``...nous avons pu constater la formation d'un nouvel isotope l\'eger du s\'el\'enium, de p\'eriode 5$\pm$2~mn, \'emetteur $\beta$, pr\'esentant une raie $\gamma$ vers 160~keV dont la masse est sans doute 71, la masse 69 n'\'etant toutefois pas exclue.'' [...we observed the formation of a new light selenium isotope, a $\beta$-emitter with a period of 5$\pm$2~min period and a 160~keV $\gamma$-ray whose mass is probably 71, however, the mass 69 cannot be excluded.] This half-life agrees with the presently accepted value of 4.74.(5)~min. A 44~min. half-life had previously been incorrectly assigned to $^{71}$Se \cite{1948Hop01}.

\subsection{$^{72}$Se}\vspace{0.0cm}
Hopkins and Cunningham first observed $^{72}$Se in 1948, in the paper ``Nuclear Reactions of Arsenic with 190-Mev Deuterons'' \cite{1948Hop01}.   190 MeV deuterons from the Berkeley 184-inch cyclotron bombarded a pure $^{75}$As target and $^{72}$Se was identified by measuring the activity with argon-filled Geiger-M\"uller counting tubes following chemical separation. ``Several new radioactivities have been observed. Mass assignments have been made by demonstration of the following decay chains. Se$^{72} \rightarrow$ As$^{72} \rightarrow$ Ge$^{72}$.''  The observed half-life of 9.5~d agrees with the accepted value of 8.40(8)~d.

\subsection{$^{73}$Se}\vspace{0.0cm}
In the 1948 paper ``Artificially Radioactive $^{73}$Se and $^{75}$Se'' Cowart et al. first identified $^{73}$Se \cite{1948Cow01}. Germanium targets were bombarded with $\alpha$ particles from the Ohio State cyclotron. Half-lives of X-rays, $\gamma$-rays and $\beta$-particles were measured to identify $^{73}$Se. ``Because of the positron emission, the 7.1-hour activity could be placed in mass 73 or 75 of selenium. An attempt was made to produce this short period by bombardment of As$^{75}$ with deuterons. Since this period was not found as a result of such bombardments, mass 73 is most probable.'' This assignment was confirmed by $\alpha$ bombardment of enriched $^{70}$Ge targets. The measured half-life of 7.1~h agrees with the accepted value of 7.15(8)~h.

\subsection{$^{74}$Se}\vspace{0.0cm}
Aston discovered $^{74}$Se in 1922 as reported in ``The Isotopes of Selenium and some other Elements'' \cite{1922Ast03}.  Selenium was vaporized in a discharge tube to obtain suitable spectra with the Cavendish mass spectrometer. ``The interpretation of these is quite simple and definite, so that the results may be stated with every confidence. Selenium consists of six isotopes, giving lines at 74(f), 76(c), 77(e), 78(b), 80(a), 82(d). The line at 74 is extremely faint. The intensities of the lines are in the order indicated by the letters, and agree well enough with the chemical atomic weight 79.2.''

\subsection{$^{75}$Se}\vspace{0.0cm}
Friedlander et al. reported the first observation of $^{75}$Se in the 1947 paper ``Evidence for, and Cross Section of 115 Day $^{75}$Se'' \cite{1947Fri01}. Selenium was irradiated in the Argonne pile and $^{75}$Ar was produced by neutron capture reactions. Samples mounted on scotch tape foils were counted with a Geiger counter. ``Evidence has been given to show that the 115-day activity produced in selenium by thermal neutrons is due to Se$^{75}$ which decays by K electron capture to As$^{75}$, accompanied by a 0.4 Mev $\gamma$-ray.'' The half-life of 115(5)~d agrees with the presently accepted value of 119.779(4)~d. The 1944 table of isotopes \cite{1944Sea01} had listed half-lives of 48~d and 160~d which were based on a private communication and a conference abstract, respectively. In addition, the headquarters of the Manhattan project listed $^{75}$Se as an available isotope with a half-life of 125~d \cite{1946ARI01}, which was based on a classified report \cite{1944Ges01}.

\subsection{$^{76-78}$Se}\vspace{0.0cm}
Aston discovered $^{76}$Se, $^{77}$Se, and $^{78}$Se in 1922 as reported in ``The Isotopes of Selenium and some other Elements'' \cite{1922Ast03}.  Selenium was vaporized in a discharge tube to obtain suitable spectra with the Cavendish mass spectrometer. ``The interpretation of these is quite simple and definite, so that the results may be stated with every confidence. Selenium consists of six isotopes, giving lines at 74(f), 76(c), 77(e), 78(b), 80(a), 82(d). The line at 74 is extremely faint. The intensities of the lines are in the order indicated by the letters, and agree well enough with the chemical atomic weight 79.2.''

\subsection{$^{79}$Se}\vspace{0.0cm}
$^{79}$Se was identified in the 1950 paper ``Bestimmung der Massenzahl der 3,9-min-Aktivit\"at des Selens'' by Flammersfeld and Herr \cite{1950Fla01}. A BrH solution was irradiated by neutrons produced by Li+d and Be+d reactions with 1.4 MeV deuterons at Mainz, Germany. The selenium activity was measured with a Geiger M\"uller counter following chemical separation. ``Die geschilderten Tatsachen scheinen hinreichend, die Zuordnung der 3,9-min-Aktivit\"at zur Massenzahl 79 zu gestatten, so da\ss\ diese also als $^{79}$Se$^*$ zu betrachten ist.'' [These facts seem sufficient to assign the 3.9~min. activity to mass 79, therefore it would correspond to $^{79}$Se$^*$.] The half-life agrees with the accepted value of 3.92(1)~min. for the isomeric state. The half-life had been observed before, however, without a unique mass assignment. It was not possible to distinguish between $^{79}$Se and $^{81}$Se \cite{1950Fla02}. The first observation of the ground state in a refereed journal was a measurement determining the spin \cite{1952Har01}.

\subsection{$^{80}$Se}\vspace{0.0cm}
Aston discovered $^{80}$Se in 1922 as reported in ``The Isotopes of Selenium and some other Elements'' \cite{1922Ast03}.  Selenium was vaporized in a discharge tube to obtain suitable spectra with the Cavendish mass spectrometer. ``The interpretation of these is quite simple and definite, so that the results may be stated with every confidence. Selenium consists of six isotopes, giving lines at 74(f), 76(c), 77(e), 78(b), 80(a), 82(d). The line at 74 is extremely faint. The intensities of the lines are in the order indicated by the letters, and agree well enough with the chemical atomic weight 79.2.''

\subsection{$^{81}$Se}\vspace{0.0cm}
The first identification of $^{81}$Se was reported by W\"affler and Hirzel in ``Relative Wirkungsquerschnitte f\"ur den ($\gamma$,n)-Prozess mit der Lithium-Gamma-Strahlung (Quantenenergie h$\nu$ = 17,5 MeV)'' \cite{1948Waf01}. $^{81}$Se was produced with the photo-nuclear reaction with lithium $\gamma$-rays on $^{82}$Se targets and identified by measuring the resulting activities. ``Der Nachweis des ($\gamma$,n)-Prozesses erfolgte jeweils vermittelst der Radioaktivit\"at des Endkerns.'' [The identification of the ($\gamma$,n) process was made by the activity of the final nucleus.] Many isotopes were measured and the results are not discussed individually but summarized in a table. For $^{81}$Se half-lives of 56.5 and 13.6~min. were listed, where the former corresponds to an isomeric state. These values are close to the presently accepted values of 57.28(2) and 18.45(12)~min., respectively. Previously, half-lives of 57~min. \cite{1937Sne01,1940Lan01} and 19~min. \cite{1940Lan01} were reported but only assigned to either $^{79}$Se or $^{81}$Se. In addition, half-lives of 57 and 17~min. could only be assigned to selenium masses lighter than 82 \cite{1939Bot01}.

\subsection{$^{82}$Se}\vspace{0.0cm}
Aston discovered $^{82}$Se in 1922 as reported in ``The Isotopes of Selenium and some other Elements'' \cite{1922Ast03}.  Selenium was vaporized in a discharge tube to obtain suitable spectra with the Cavendish mass spectrometer. ``The interpretation of these is quite simple and definite, so that the results may be stated with every confidence. Selenium consists of six isotopes, giving lines at 74(f), 76(c), 77(e), 78(b), 80(a), 82(d). The line at 74 is extremely faint. The intensities of the lines are in the order indicated by the letters, and agree well enough with the chemical atomic weight 79.2.''

\subsection{$^{83}$Se}\vspace{0.0cm}
Snell discovered $^{83}$Se in 1937 as reported in ``The Radioactive Isotopes of Bromine: Isomeric Forms of Bromine 80'' \cite{1937Sne01}. The Berkeley cyclotron was used to bombard selenium targets with 5.5 MeV deuterons and $^{83}$Se was produced in the reaction $^{82}$Se(d,p). The activities were measured with a quartz-fiber electroscope following chemical separation. The observation of $^{83}$Se was discussed in the context of the observation of $^{83}$Br which was proposed to be produced in the decay of $^{83}$Se. ``...By growth from a hitherto unknown selenium 83; i.e., the bromine 83 would be the second member in a pair of successive $\beta$-emitters, the reactions being Se$^{82}$(D,p)Se$^{83}$, Se$^{83} \rightarrow$ Br$^{83}$ + e$^-$, Br$^{83} \rightarrow$ Kr$^{83}$ + e$^-$. There was evidence for the second process in the decay curves of the selenium fractions obtained in the experiments, and it appeared that the selenium 83 had a half-life of 17$\pm$5 minutes.'' This half-life is consistent with the accepted half-life of 22.3(3)~min.

\subsection{$^{84,85}$Se}\vspace{0.0cm}
In the 1960 paper ``Short-lived Bromine and Selenium Nuclides From Fission'' Sattizahn et al. reported the discovery of $^{84}$Se and $^{85}$Se \cite{1960Sat01}. The isotopes were produced by irradiation of 93\% $^{235}$U in the Los Alamos Water Boiler. Decay curves were measured with methane-flow proportional counters following chemical separation. ``The half-lives of $^{84}$Se and $^{85}$Se were determined by periodic extraction and measurement of the daughter 31.7~min. and 3.0~min. bromine activities which grow from fission-product selenium.'' The observed half-life of 3.3(3)~min. for $^{84}$Se is included in the calculation of the currently accepted average value of 3.1(1)~min. and the half-life of 39(4)~s for $^{85}$Se agrees with the present value of 31.7(9)~s. The 1948 Table of Isotopes \cite{1948Sea01} quotes half-lives of 2.5~min \cite{1946Gle01} and $<$10~min. \cite{1944Edw01} for $^{84}$Se attributed to results from the Manhattan project. Both were classified reports, and only the latter one quoting an upper limit for the half-life was included in the unclassified publication of the Plutonium Project Records \cite{1951Edw01}.

\subsection{$^{86}$Se}\vspace{0.0cm}
Tamai et al. observed $^{86}$Se in 1973 as reported in ``Gamma-Ray Energies of Se-85 and Se-86'' \cite{1973Tam01}. A 90\% enriched uranyl nitrate solution was irradiated with thermal neutrons in the Kyoto University Reactor. Gamma-ray spectra were measured with a Ge(Li) detector following chemical separation. In the text of the article the assignment is not convincing: ``The photopeaks with half-life of 15~s are not necessarily due to the $\gamma$-ray from $^{86}$Se because the half-life of $^{88}$Br (17.5~s) is nearly the same as that of $^{86}$Se (16~s). Accordingly, from the half-lives alone $^{86}$Se cannot be distinguished from $^{86}$Br.'' However, the figures and the last table clearly assign $\gamma$-rays to $^{86}$Se and the current data evaluations concur with this assignment for the lowest three energies. The 16~s half-life mentioned in the quote had initially been assigned to $^{87}$Se \cite{1960Sat01}. It is not clear when this assignment was changed. While the 1967 Table of Isotopes \cite{1967Led01} still listed it for $^{87}$Se, the 1969 review article by Herrmann and Denschlag \cite{1969Her01} assigned the 16~s half-life to $^{86}$Se.

\subsection{$^{87}$Se}\vspace{0.0cm}
Tomlinson discovered $^{87}$Se in the 1968 paper ``Delayed Neutron Precursors-III Selenium-87'' \cite{1968Tom01}. $^{87}$Se was produced by neutron irradiation in the LIDO reactor at Harwell, England. The half-life of $^{87}$Se was measured ``...by separating selenium in the trap at various times after irradiation and counting the neutrons from 55 sec $^{87}$Br grown in from $^{87}$Se... A least-squares fit gave a half-life (with standard deviation) of 5.8 $\pm$ 0.5 sec for $^{87}$Se,...''. This half-life agrees with the presently accepted value of 5.50(12)~s. Previously, a 16~s half-life - most likely $^{86}$Se - had been assigned incorrectly to $^{87}$Se \cite{1960Sat01}.

\subsection{$^{88}$Se}\vspace{0.0cm}
In the 1970 paper ``Identification of $^{88}$Se and search for Delayed Neutron Emission from $^{87}$Se and $^{88}$Se'', del Marmol and Perricos reported the discovery of $^{88}$Se \cite{1970Mar01}. $^{235}$U was irradiated with thermal neutrons in the core of the BR1 reactor in Mol, Belgium. The half-life was determined from the neutron activities of the bromine daughter following chemical separation. ``From a set of experiments with a 5.05 sec irradiation  time... a first estimate of about 1.7 sec was obtained for the half-life of $^{88}$Se... Although the latter measurements [(1.1(1)~s for 3.05 sec irradiation times)] are believed to be more reliable than those for the 5.05 sec irradiation time a conservative estimate of 1.3$\pm$0.3 sec was taken for the half-life of $^{88}$Se.'' This half-life is in agreement with the accepted value of 1.53(6)~s.

\subsection{$^{89}$Se}\vspace{0.0cm}
Tomlinson and Hurdus observed $^{89}$Se as reported in the 1971 paper ``Delayed Neutron Precursors-IV $^{87}$Se, $^{88}$Se and $^{89}$Se Half-lives, Neutron Emission Probabilities and Fission Yields'' \cite{1971Tom01}. $^{89}$Se was produced by neutron irradiation in the LIDO reactor at Harwell, England. Delayed neutron emission was measured following rapid chemical separation. ``The following data have been obtained:... $^{89}$Se: half-life, 0.41 $\pm$ 0.04 sec;... $^{89}$Se is a new nuclide, identified for the first time in this work.'' This half-life corresponds to the presently accepted value.

\subsection{$^{90}$Se}\vspace{0.0cm}
Bernas {\it{et al.}} reported the discovery of $^{90}$Se in the 1994 paper ``Projectile Fission at Relativistic Velocities: A Novel and Powerful Source of Neutron-Rich Isotopes Well Suited for In-Flight Isotopic Separation'' \cite{1994Ber01}. The isotope was produced using projectile fission of $^{238}$U at 750 MeV/nucleon on a lead target at GSI, Germany. ``Forward emitted fragments from $^{80}$Zn up to $^{155}$Ce were analyzed with the Fragment Separator (FRS) and unambiguously identified by their energy-loss and time-of-flight.'' The experiment yielded 409 individual counts of $^{90}$Se.

\subsection{$^{91}$Se}\vspace{0.0cm}
In the 1975 publication ``The P$_n$ Values of the $^{235}$U(n$_{th}$,f) Produced Precursors in the Mass Chains 90, 91, 93-95, 99, 134 and 137-139'' Asghar et al. reported the first observation of $^{91}$Se \cite{1975Asg01}. The isotope was produced by thermal neutron fission of $^{235}$U in the Grenoble high flux reactor and identified with the mass separator Lohengrin. ``The present work led to: (i) three new periods corresponding to the new isotopes of selenium ($^{91}$Se, T$_{1/2}$ = 0.27$\pm$0.05~sec), strontium ($^{99}$Sr, T$_{1/2}$ = 0.6$\pm$0.2~sec) and tellurium ($^{138}$Te, T$_{1/2}$ = 1.3$\pm$0.3~sec)...'' The half-life corresponds to the presently accepted value.

\subsection{$^{92-94}$Se}\vspace{0.0cm}
$^{92-94}$Se were discovered by Bernas et al. in 1997 reported in ``Discovery and Cross-Section Measurement of 58 New Fission Products in Projectile-Fission of 750$\cdot$AMeV $^{238}$U'' \cite{1997Ber01}. The experiment was performed using projectile fission of $^{238}$U at 750~MeV/nucleon on a beryllium target at GSI in Germany. ``Fission fragments were separated using the fragment separator FRS tuned in an achromatic mode and identified by event-by-event measurements of $\Delta$E-B$\rho$-ToF and trajectory.'' During the experiment, individual counts for $^{92}$Se (1061), $^{93}$Se (117), and $^{94}$Se (31) were recorded.

\subsection{$^{95}$Se}\vspace{0.0cm}
The discovery of $^{95}$Se was reported in the 2010 article ``Identification of 45 New Neutron-Rich Isotopes Produced by In-Flight Fission of a $^{238}$U Beam at 345 MeV/nucleon,'' by Ohnishi et al. \cite{2010Ohn01}. The experiment was performed at the RI Beam Factory at RIKEN, where the new isotopes were created by in-flight fission of a 345 MeV/nucleon $^{238}$U beam on a beryllium target. $^{95}$Se was separated and identified with the BigRIPS superconducting in-flight separator. The list of new isotopes discovered in this study are summarized in a table. Fifteen individual counts of $^{95}$Se were recorded.

\section{Discovery of $^{70-98}$Br}

Twenty-nine bromine isotopes from A = $70-98$ have been discovered so far; these include 2 stable, 10 proton-rich and 17 neutron-rich isotopes.  According to the HFB-14 model \cite{2007Gor01}, $^{116}$Br should be the last odd-odd particle stable neutron-rich nucleus while the odd-even particle stable neutron-rich nuclei should continue through $^{119}$Br. The proton dripline has been reached and no more long-lived isotopes are expected to exist because $^{69}$Br has been shown to be unbound with a half-life of less than 100~ns \cite{1995Bla01}. A previous report of the observation of $^{69}$Br \cite{1991Moh01} was evidently incorrect. About 20 isotopes have yet to be discovered corresponding to 41\% of all possible bromine isotopes.

\begin{figure}
	\centering
	 \includegraphics[scale=.5]{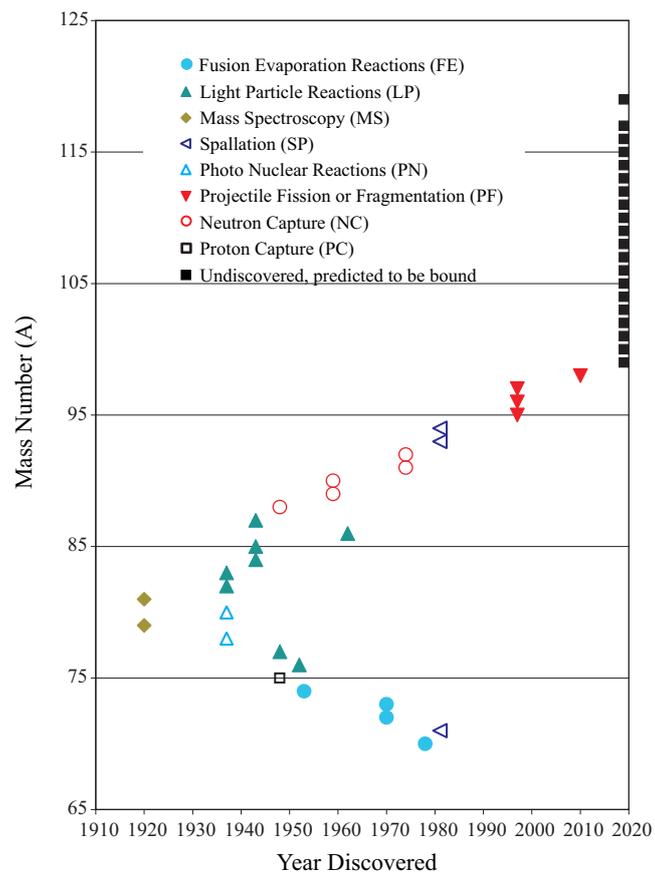}
	\caption{Bromine isotopes as a function of time when they were discovered. The different production methods are indicated. The solid black squares on the right hand side of the plot are isotopes predicted to be bound by the HFB-14 model.}
\label{f:year-br}
\end{figure}

Figure \ref{f:year-br} summarizes the year of first discovery for all bromine isotopes identified by the method of discovery. The range of isotopes predicted to exist is indicated on the right side of the figure. The radioactive bromine isotopes were produced using heavy-ion fusion-evaporation reactions (FE), light-particle reactions (LP), neutron induced fission (NF), proton capture reactions (PC), photo-nuclear reactions (PN), spallation (SP), and projectile fragmentation of fission (PF). The stable isotopes were identified using mass spectroscopy (MS). Heavy ions are all nuclei with an atomic mass larger than A=4 \cite{1977Gru01}. Light particles also include neutrons produced by accelerators. In the following, the discovery of each bromine isotope is discussed in detail and a summary is presented in Table 1.

\subsection{$^{70}$Br}\vspace{0.0cm}
Alburger discovered $^{70}$Br in 1978 which he announced in ``Half Lives of $^{62}$Ga, $^{66}$As, and $^{70}$Br'' \cite{1978Alb01}. A 44 MeV $^{14}$N beam accelerated by the Brookhaven MP Tandem Van de Graaff bombarded an enriched $^{58}$Ni foil. $^{70}$Br was produced in the fusion-evaporation reaction $^{56}$Ni($^{14}$N,2n)$^{70}$Br and identified by measuring the $\beta$-ray decay curve following activation with an NE102 scintillation detector. ``Six runs were made on $^{70}$Br totaling about 50 hours of data taking. By analysis of the results... the value adopted for the half-life of $^{70}$Br is 80.2$\pm$0.8 ms. Although firm proof is lacking that the observed activity is in fact $^{70}$Br, the assignment is highly probable because of the agreement of both the measured yield and the half-life with calculated estimates.'' The recorded half-life of 80.2(8)~ms is part of the calculation of the currently accepted average of 79.1(8)~ms.

\subsection{$^{71}$Br}\vspace{0.0cm}
Vosicki et al. were the first to observe $^{71}$Br in 1981 reported in the paper ``Intense Beams of Radioactive Halogens produced by Means of Surface Ionization'' \cite{1981Vos01}. $^{71}$Br was produced by bombarding a niobium-powder and a UC-graphite target with 600 MeV protons at the CERN ISOLDE facility. The isotope was separated and identified with a negative surface-ionization source. ``During periods of several weeks the source has been used for a number of on-line experiments and has allowed the identification of a series of new nuclei. Yields and half-lives of $^{42,43}$Cl and $^{70,71,72m,76m,93,94}$Br are reported.'' The measured half-life for $^{71}$Br was 21.5(5)~s which agrees with the accepted half-life value of 21.4(6)~s.

\subsection{$^{72}$Br}\vspace{0.0cm}
Nolte et al. discovered $^{72}$Br in 1970 as described in the paper ``Neutron Deficient Even-Even Isotopes of Kr, Se, Ge and Pseudo SU$_3$ coupling'' \cite{1970Nol01}. Beams of 42-57 MeV $^{16}$O were accelerated by the Heidelberg EN and MP tandem Van de Graaff accelerators and bombarded enriched $^{58}$Ni targets. Prompt and delayed $\gamma$-ray spectra were measured with two Ge(Li) detectors. ``The new isotope $^{72}$Br was found through the reaction $^{58}$Ni($^{16}$O,pn)$^{72}$Br. It decays with a half life of approximately 1.7 min to the first and second 2$^+$ states in $^{72}$Se.'' This half-life agrees with the accepted half-life of 78.6(24)~s. It should be mentioned that Doron and Blann reported the observation of $^{72}$Br only a month later \cite{1971Dor01}.

\subsection{$^{73}$Br}\vspace{0.0cm}
The first observation of $^{73}$Br was reported in 1969 by Murray et al. in ``The decay of $^{73}$Br'' \cite{1970Mur01}. The Manchester Hilac accelerated a 64 MeV $^{16}$O beam which bombarded a cobalt target to form $^{73}$Br in the fusion-evaporation reaction $^{59}$Co($^{16}$O,2n). Gamma-ray spectra were measured with a Ge(Li) following chemical separation. ``There was no evidence for any initial build-up in the $\gamma$-ray yields and this, together with the fact that chemical separation took place 2 min after the end of the irradiation, can only be interpreted on the basis that the 3.3$\pm$0.3 min activity is associated with the decay of a bromine isotope... The initial build-up in the decay curves is consistent with a 3$\pm$1 min activity feeding the $^{73}$Se 40 min level. We thus assign the 3.3$\pm$0.3 min activity to $^{73}$Br.'' This half-life agrees with the currently accepted value of 3.4(2)~m. A 4~min. activity had previously been observed, however, no mass assignment was possible \cite{1953Hol01,1960But01}.

\subsection{$^{74}$Br}\vspace{0.0cm}
$^{74}$Br was discovered by Hollander in 1953 as described in ``Bromine Isotopes Produced by Carbon-Ion Bombardment of Copper'' \cite{1953Hol01}. An approximately 90 MeV $^{12}$C beam from the Berkeley 60-inch cyclotron bombarded copper foils and $^{74}$Br was formed in the fusion-evaporation reaction $^{63}$Cu($^{12}$C,n). Decay curves were recorded following chemical separation. Targets with enriched $^{63}$Cu and $^{65}$Cu were used and the results indicated ``that the 36-minute activity has been made in Cu$^{63}$ by a reaction with a very low cross section, and hence may be assigned tentatively to Br$^{74}$, by the Cu$^{63}$(C,n)Br$^{74}$ reaction.'' This half-life is between the half-life of the ground-state (25.4(3)~min.) and an isomeric state (46(2)~min.).

\subsection{$^{75}$Br}\vspace{0.0cm}
Woodward et al. described the first observation of $^{75}$Br in the 1948 paper ``Radioactive Br Isotopes'' \cite{1948Woo02}. Deuterons and protons from the Ohio State cyclotron bombarded enriched $^{74}$Se targets. Gamma- and $\beta$-rays were measured with a Wulf electrometer attached to a Freon-filled ionization chamber. ``In the decay curve of activity from Se$^{74}$+p bombardment, a 125$\pm$5-day half-life appeared which was presumed to be that of Se$^{75}$. The ratio of reaction cross sections for production of the 1.7-hour Br activity compared to the 127-day Se$^{75}$ activity by proton bombardment was determined to be approximately one. This indicated that the 1.7-hour Br activity decays into the 127-day Se$^{75}$ period. From this data, assignment of the 1.7-hour activity was made to Br$^{75}$ as a Se$^{74}$(p,$\gamma$) reaction.'' This half-life is consistent with the currently accepted value of 96.7(13)~min.

\subsection{$^{76}$Br}\vspace{0.0cm}
In 1951 $^{76}$Br was discovered by Fultz and Pool as reported in ``Radioisotopes of Bromine'' \cite{1952Ful01}. A 7.3 MeV proton beam bombarded selenium metal enriched with $^{76}$Se at Ohio State University. The activity was measured with a Wulf electrometer filled with freon. ``In addition to the well-known bromine activities of 4.4-hour Br$^{80}$, 36-hour Br$^{82}$, and 2.4-day Br$^{77}$, a new activity of 17.2 hours was observed... Thus it is seen that within limits of experimental error, no evidence for a (p,$\gamma$) reaction exists. Hence it is concluded that the 17.2-hour activity arises from Se$^{76}$ by (p,n) reaction only, and is therefore assigned to $^{76}$Br.'' This half-life is consistent with the presently adopted value of 16.2(2)~h.

\subsection{$^{77}$Br}\vspace{0.0cm}
Woodward et al. described the first observation of $^{75}$Br in the 1948 paper ``Radioactive Br Isotopes'' \cite{1948Woo02}. Deuterons and $\alpha$-particles from the Ohio State cyclotron bombarded enriched $^{74}$Se and $^{76}$Se targets. Gamma- and $\beta$-rays were measured with a Wulf electrometer attached to a Freon-filled ionization chamber. A 2.4-day activity was observed with $\alpha$-particles on $^{74}$Se and with deuterons on $^{76}$Se. Following the latter observation it was concluded that ``Since bombardment of Se with alpha-particles had previously located the activity as either Kr$^{77}$ or Br$^{77}$, assignment of the activity was made to Br$^{77}$.'' This half-life is consistent with the currently accepted value of 57.036(6)~h.

\subsection{$^{78}$Br}\vspace{0.0cm}
The first observation of $^{78}$Br was reported in the 1937 paper ``Herstellung neuer Isotope durch Kernphotoeffekt'' by Bothe and Gentner \cite{1937Bot01}. $^{78}$Br was produced in photo-nuclear reactions with lithium $\gamma$-rays. ``Brom (Br$^{79}$ + Br$^{81}$) zeigte au\ss er dem schon mitgeteilten 18-min-Abfall des Br$^{80}$ noch einen Abfall mit 3,5 min Halbwertzeit; als Tr\"ager wurde wiederum Brom chemisch nachgewiesen, es handelt sich also um Br$^{78}$.'' [Brom (Br$^{79}$ + Br$^{81}$) exhibited in addition to the already published 18-min decay of $^{80}$Br another decay of 3.5~min; again it was chemically shown to be bromine, therefore it has to be $^{78}$Br.] The measured half-life is somewhat smaller (within a factor of two) than the presently accepted half-life of 6.45(4)~min. More accurate half-lives of 5~min. \cite{1937Cha01,1937Hey01} and 6.4~min. \cite{1937Sne01} were reported later in the same year. A 36-h half-life had previously been incorrectly assigned to $^{78}$Br \cite{1935Kur01}.

\subsection{$^{79}$Br}\vspace{0.0cm}

Aston discovered $^{79}$Br in 1920 as reported in ``The constitution of the elements'' \cite{1920Ast02}. The isotope was identified by measuring its mass spectra and no further experimental details were given. ``Bromine (atomic weight 79.92) is particularly interesting, for, although its chemical atomic weight is so nearly 80, it is actually composed of approximately equal parts of isotopes 79 and 81.''

\subsection{$^{80}$Br}\vspace{0.0cm}

In the 1937 article ``K\"unstliche Radioaktivit\"at durch $\gamma$-Strahlen'' Bothe and Gentner identified $^{80}$Br \cite{1937Bot02}. 17.5 MeV $\gamma$-rays were produced by the bombardment of 450 keV protons on lithium irradiated bromine targets and $^{80}$Br was produced by photonuclear reactions. ``Als aktives Produkt wurde chemisch Brom nachgewiesen. Die Halbwertszeit deckt sich innerhalb der Me\ss genauigkeit mit derjenigen des einen bekannten Bromisotops, n\"amlich 18 min. Dieses Isotop kann auch durch Anlagerung langsamer Neutronen aus gew\"ohnlichem Brom (Br$^{79}$ + Br$^{81}$) gebildet werden. Daher kann diese Halbwertszeit jetzt dem Br$^{80}$ zugeordnet werden, gem\"a\ss\  Br$^{81}$($\gamma$,n)Br$^{80}$ bzw. Br$^{79}$(n,$\gamma$)Br$^{80}$. [The activity was chemically identified as bromine. The half-life agrees within the uncertainties with the known 18-min. bromine isotope. This isotope can also be produced by slow neutron capture on natural bromine (Br$^{79}$ + Br$^{81}$). Thus, this half-life can now be assigned to Br$^{80}$, according to the reactions Br$^{81}$($\gamma$,n)Br$^{80}$ and Br$^{79}$(n,$\gamma$)Br$^{80}$.] This half-life agrees with the presently accepted value of 17.68(2)~min. Amaldi et al. \cite{1935Ama01} had observed the 18~min. half-life in addition to the 4.2~h isomer earlier but could not make a unique mass assignment.

\subsection{$^{81}$Br}\vspace{0.0cm}

Aston discovered $^{81}$Br in 1920 as reported in ``The constitution of the elements'' \cite{1920Ast02}. The isotope was identified by measuring its mass spectra and no further experimental details were given. ``Bromine (atomic weight 79.92) is particularly interesting, for, although its chemical atomic weight is so nearly 80, it is actually composed of approximately equal parts of isotopes 79 and 81.''

\subsection{$^{82,83}$Br}\vspace{0.0cm}

Snell discovered $^{82}$Br and $^{83}$Br in 1937 which he announced in ``The Radioactive Isotopes of Bromine: Isomeric Forms of Bromine 80'' \cite{1937Sne01}. The Berkeley cyclotron was used to accelerate deuterons and $\alpha$-particles to produce neutrons which were then used to irradiate samples of arsenic, selenium and bromine. A variety of reactions were used to assign the three previously observed half-lives of 18~min., 4.5~h and 35~h to the correct bromine isotopes. ``The experiments led to the conclusion that the only satisfactory explanation of the occurrence of the three bromine activities lies in ascribing two activities to bromine 80. These are the 18-min. and the 4.5-hr. activities; the 35-hr activity appears to belong to bromine 82.'' $^{83}$Br was proposed to be produced in the decay of $^{83}$Se. ``...By growth from a hitherto unknown selenium 83; i.e., the bromine 83 would be the second member in a pair of successive $\beta$-emitters, the reactions being Se$^{82}$(D,p)Se$^{83}$, Se$^{83} \rightarrow$ Br$^{83}$ + e$^-$, Br$^{83} \rightarrow$ Kr$^{83}$ + e$^-$.'' The measured half-lives of 33.9(3)~h and 2.54(10)~h are consistent with the presently accepted value of 35.28(7)~h and 2.40(2)~h for $^{82}$Br and $^{83}$Br, respectively. A 140(10)~min. half-life had been reported without a mass assignment \cite{1940Str01}. Previously a 36~h half-life had been incorrectly assigned to $^{78}$Br \cite{1935Kur01}.

\subsection{$^{84,85}$Br}\vspace{0.0cm}

In 1943, Born and Seelmann-Eggebert were the first to identify $^{84}$Br and $^{85}$Br in their paper ``\"Uber die Identifizierung einiger Uranspaltprodukte mit entsprechenden durch (n$\alpha$)- und (np)-Prozesse erhaltenen Isotopen'' \cite{1943Bor01}. Rubidium and strontium salts were irradiated with neutrons from the high-voltage facility of the Kaiser Wilhelm Institut f\"ur Physik in Berlin, Germany and decay curves following chemical separation were measured. ``Das 30-Min.-Brom aus Uran ist identisch mit einem aus Rubidium durch (n$\alpha$)-Proze\ss\ erhaltenem Brom. Auf Grund dieser Ergebnisse k\"onnen folgende Reihen aufgestellt werden: $^{84}$Br (30~Min.) $\rightarrow ^{84}$Kr stab., $^{85}$Br (3~Min.) $\rightarrow ^{85}$Kr (4,6~Std.) $\rightarrow ^{85}$Rb stab., $^{87}$Br (50~Sek.) $\rightarrow ^{87}$Kr (75~Min.) $\rightarrow ^{87}$Rb (6,3 $\cdot$ 10$^{10}$ Jahre) $\rightarrow ^{87}$Sr stab.'' [The 30~min. bromine from uranium fission is identical to the bromine produced by the (n,$\alpha$) process from rubidium. Based on these results we can determine the following decay sequences: $^{84}$Br (30~min.) $\rightarrow ^{84}$Kr stable, $^{85}$Br (3~min.) $\rightarrow ^{85}$Kr (4,6~h) $\rightarrow ^{85}$Rb stable, $^{87}$Br (50~s) $\rightarrow ^{87}$Kr (75~min.) $\rightarrow ^{87}$Rb (6.3 $\cdot$ 10$^{10}$~y) $\rightarrow ^{87}$Sr stable.] The measured half-lives of 30~min. for $^{84}$Br and 3~min. for $^{85}$Br are consistent with the presently adopted values of 31.80(8)~min. and 2.90(6)~min., respectively. The half-lives of both isotopes had been previously reported without mass assignments, 35~min. \cite{1939Hah01}, 40~min. \cite{1939Dod01}, and 30(5)~min. \cite{1940Str01} for $^{84}$Br and 3.0(5)~min. \cite{1940Str01} for $^{85}$Br.

\subsection{$^{86}$Br}\vspace{0.0cm}

Stehney and Steinberg discovered $^{86}$Br in 1962 as reported in ``Br$^{86}$-A New Nuclide'' \cite{1962Ste01}. Neutrons with a maximum energy of 25 MeV were produced by bombarding 21.5 MeV deuterons from the Argonne 60-in. cyclotron on a beryllium target. $^{86}$Br was then produced in the (n,p) charge exchange reaction on an enriched $^{86}$Kr target. The half-life was determined by $\beta$- and $\gamma$-ray spectroscopy following chemical separation. ``The assignment of the 54-sec activity to Br$^{86}$ produced by the (n,p) reaction on Kr$^{86}$ is made on the basis of the following argument: (1) It cannot be Br$^{87}$, since the formation of Br$^{87}$ by neutron bombardment of Kr would require double neutron capture. (2) It is a new nuclide, differing in half-life and decay energy from other bromine products of the neutron bombardment of krypton. (3) The Q{$_\beta$} observed is consistent with that expected for Br$^{86}$. (4) It is probably not an isomer of Br$^{85}$ since isomerism in this region is very unlikely.'' The measured half-life of 54(2)~s agrees with the currently accepted half-life of 55.1(4)~s.

\subsection{$^{87}$Br}\vspace{0.0cm}

In 1943, Born and Seelmann-Eggebert were the first to identify $^{87}$Br in their paper ``\"Uber die Identifizierung einiger Uranspaltprodukte mit entsprechenden durch (n$\alpha$)- und (np)-Prozesse erhaltenen Isotopen'' \cite{1943Bor01}. Rubidium and strontium salts were irradiated with neutrons from the high-voltage facility of the Kaiser Wilhelm Institut f\"ur Physik in Berlin, Germany and decay curves following chemical separation were measured. ``Das 30-Min.-Brom aus Uran ist identisch mit einem aus Rubidium durch (n$\alpha$)-Proze\ss\ erhaltenem Brom. Auf Grund dieser Ergebnisse k\"onnen folgende Reihen aufgestellt werden: $^{84}$Br (30~Min.) $\rightarrow ^{84}$Kr stab., $^{85}$Br (3~Min.) $\rightarrow ^{85}$Kr (4,6~Std.) $\rightarrow ^{85}$Rb stab., $^{87}$Br (50~Sek.) $\rightarrow ^{87}$Kr (75~Min.) $\rightarrow ^{87}$Rb (6,3 $\cdot$ 10$^{10}$ Jahre) $\rightarrow ^{87}$Sr stab.'' [The 30~min. bromine from uranium fission is identical to the bromine produced by the (n,$\alpha$) process from rubidium. Based on these results we can determine the following decay sequences: $^{84}$Br (30~min.) $\rightarrow ^{84}$Kr stable, $^{85}$Br (3~min.) $\rightarrow ^{85}$Kr (4,6~h) $\rightarrow ^{85}$Rb stable, $^{87}$Br (50~s) $\rightarrow ^{87}$Kr (75~min.) $\rightarrow ^{87}$Rb (6.3 $\cdot$ 10$^{10}$~y) $\rightarrow ^{87}$Sr stable.] The measured half-life of 50~s is consistent with the presently accepted value of 55.65(13)~s. The half-life 50(10)~s \cite{1940Str01} had been previously reported without a unique mass assignment.

\subsection{$^{88}$Br}\vspace{0.0cm}

The discovery of $^{88}$Br was reported by Sugarman in the 1948 publication ``Short-Lived Halogen Fission Products'' \cite{1949Sug01}. $^{88}$Br was produced in the fission of irradiated uranyl nitrate. Activities were measured following chemical separation. ``After corrections were made for the chemical yield of Rb, for decay of the 2.8-hr. Kr before growth of 17.8-min. Rb started, and for growth of 17.8-min. Rb during the standing period (1 hour), the Kr activities so derived were extrapolated to the time of AgBr precipitation and normalized by the 4.5-hr. Kr normalization factors of [the Table]. Least squares analysis of the normalized activities versus the time of precipitation yielded a half-life of 15.5$\pm$0.3 sec. for Br$^{88}$.'' This half-life agrees with the presently adopted value of 16.29(6)~s.

\subsection{$^{89,90}$Br}\vspace{0.0cm}

Perlow and Stehney observed $^{89}$Br and $^{90}$Br as described in the 1958 article ``Halogen Delayed-Neutron Activities'' \cite{1959Per01}. $^{235}$U was irradiated with thermal neutrons and the activities of the bromine isotopes were measured following chemical separation. ``The 4.4-sec bromine was observed by Sugarman by neutron-counting after chemical separation, but the half-life assignment in that work was open to some question because the delayed-neutron character of Br$^{88}$ was not known at that time. The present work corroborates its existence and furnishes a good measurement of the lifetime. This isotope is frequently assigned to mass 89, although a search for mass 89 descendents following AgBr precipitation was not successful. The 1.6-sec bromine has not been observed previously. It is likely to be Br$^{90}$, which like Br$^{88}$ is also odd-odd.'' The measured half-lives of 4.4(5)~s and 1.6(6)~s agree with the presently accepted values of 4.40(3)~s and 1.910(10)~s for $^{89}$Br and $^{90}$Br, respectively. A 4.45~s \cite{1947Red01}and a 4.51~s \cite{ } half-life were reported in the fission of $^{235}$U, however, no element or isotope identification was attempted. The 4.51~s half-life was identified as bromine with possible mass numbers between 87 and 90 \cite{1947Sug01}; subsequently it was narrowed down to 89 or 90 \cite{1949Sug01}.

The quoted papers Phys. Rev. 72 (1947) 570 and Phys. Rev. 73 (1948) 111 describe a 4.45s and 4.51s half-life, respectively, in the fission of 235U. No element or isotope identification was attempted. In a separate paper (J. Chem. Phys. 15 (1947) 544) bromine is identified, however, the mass was still uncertain:"The remaining possible mass assignments of the 4.51 sec. Br are 87 to 90." We include these references now in the paper.

\subsection{$^{91,92}$Br}\vspace{0.0cm}

In 1974, Kratz and Herrmann discovered the isotopes $^{91}$Br and $^{92}$Br as published in ``Delayed-Neutron Emission from Short-Lived Br and I Isotopes'' \cite{1974Kra01}. Delayed neutron activities of chemical separated fission fragments were produced by thermal neutron fission of $^{235}$U in the Mainz Triga reactor. ``In the bromine fraction... two new activities with half-lives of 0.63 sec and 0.25 sec become visible after subtraction of the known precursors $^{87}$Br, $^{88}$Br, $^{89}$Br and $^{90}$Br and are assigned to $^{91}$Br and $^{92}$Br, respectively.'' The measured half-life of 0.63(7)~s for $^{91}$Br agrees with the presently adopted value of 0.63(7)~s and the half-life of 0.26(4)~s for $^{92}$Br is included in the calculation of the current averaged value of 0.343(15)~s.

\subsection{$^{93,94}$Br}\vspace{0.0cm}

Vosicki et al. were the first to observe $^{93}$Br and $^{94}$Br in 1981 reported in the paper ``Intense Beams of Radioactive Halogens produced by means of Surface Ionization'' \cite{1981Vos01}. The bromine isotopes were produced by bombarding a niobium-powder and a UC-graphite target with 600 MeV protons at the CERN ISOLDE facility. The isotopes were separated and identified with a negative surface-ionization source. ``During periods of several weeks the source has been used for a number of on-line experiments and has allowed the identification of a series of new nuclei. Yields and half-lives of $^{42,43}$Cl and $^{70,71,72m,76m,93,94}$Br are reported.'' The half-lives for $^{93}$Br and $^{94}$Br were actually not measured, however, a particle identification plot demonstrating the presence of the two isotopes is presented.

\subsection{$^{95-97}$Br}\vspace{0.0cm}

$^{95}$Br, $^{96}$Br, and $^{97}$Br were discovered by Bernas et al. in 1997 at GSI in Germany and reported in ``Discovery and Cross-Section Measurement of 58 New Fission Products in Projectile-Fission of 750$\cdot$AMeV $^{238}$U'' \cite{1997Ber01}. The experiment was performed using projectile fission of $^{238}$U at 750~MeV/nucleon on a beryllium target. ``Fission fragments were separated using the fragment separator FRS tuned in an achromatic mode and identified by event-by-event measurements of $\Delta$E-B$\rho$-ToF and trajectory.'' During the experiment, individual counts for $^{95}$Br (781), $^{96}$Br (69), and $^{97}$Br (8) were recorded.

\subsection{$^{98}$Br}\vspace{0.0cm}

The discovery of $^{98}$Br was reported in the 2010 article ``Identification of 45 New Neutron-Rich Isotopes Produced by In-Flight Fission of a $^{238}$U Beam at 345 MeV/nucleon,'' by Ohnishi et al. \cite{2010Ohn01}. The experiment was performed at the RI Beam Factory at RIKEN, where the new isotopes were created by in-flight fission of a 345 MeV/nucleon $^{238}$U beam on a beryllium target. $^{98}$Br was separated and identified with the BigRIPS superconducting in-flight separator. The list of new isotopes discovered in this study are summarized in a table. Eleven individual counts for $^{98}$Br were recorded.

\section{Discovery of $^{125-156}$Nd}

Thirty-one neodymium isotopes from A = 125-156 have been discovered so far; these include 7 stable, 16 proton-rich and 8 neutron-rich isotopes. $^{126}$Nd, \cite{2000Sou01} $^{157}$Nd, and $^{158}$Nd \cite{1997Ber02} have only been reported in conference proceedings, but have not yet been published in the refereed literature. According to the HFB-14 model \cite{2007Gor01}, $^{185}$Nd should be the last odd-even particle stable neutron-rich nucleus while the even-even particle stable neutron-rich nuclei should continue through $^{196}$Nd. At the proton dripline six more isotopes could be observed ($^{120}$Nd through $^{124}$Nd, and $^{126}$Nd). It is estimated that six additional nuclei beyond the proton dripline could live long enough to be observed \cite{2004Tho01}. About 47 isotopes have yet to be discovered corresponding to 60\% of all possible neodymium isotopes.

\begin{figure}
	\centering
	\includegraphics[scale=.5]{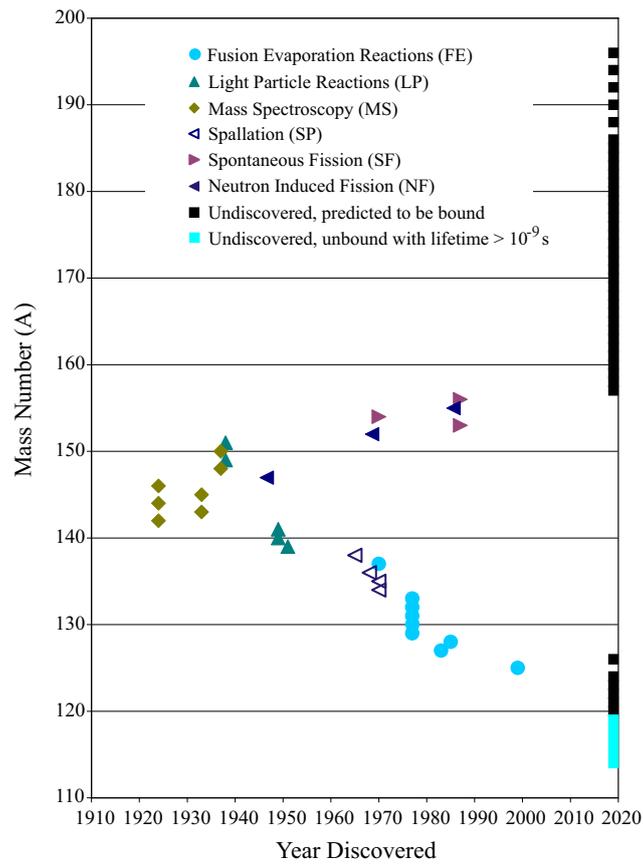}
	\caption{Neodymium isotopes as a function of time when they were discovered. The different production methods are indicated. The solid black squares on the right hand side of the plot are isotopes predicted to be bound by the HFB-14 model. On the proton-rich side the light blue squares correspond to unbound isotopes predicted to have lifetimes larger than $\sim 10^{-9}$~s.}
\label{f:year-nd}
\end{figure}

Figure \ref{f:year-nd} summarizes the year of first discovery for all neodymium isotopes identified by the method of discovery. The range of isotopes predicted to exist is indicated on the right side of the figure. The radioactive neodymium isotopes were produced using fusion evaporation reactions (FE), light-particle reactions (LP), neutron induced fission (NF), spontaneous fission (SF) and spallation reactions (SP). The stable isotopes were identified using mass spectroscopy (MS). Light particles also include neutrons produced by accelerators. In the following, the discovery of each neodymium isotope is discussed in detail and a summary is presented in Table 1.

\subsection{$^{125}$Nd}\vspace{0.0cm}

Xu et al. first identified $^{125}$Nd in 1999 and reported the results in ``New $\beta$-delayed proton precursors in the rare-earth region near the proton drip line'' \cite{1999Xu01}. A 169 MeV $^{36}$Ar beam was accelerated with the sector-focused cyclotron at the National Laboratory of Heavy-Ion Accelerator in Lanzhou, China, and bombarded an enriched $^{92}$Mo target. Proton-$\gamma$ coincidences were measured in combination with a He-jet tape transport system. ``The decay curve of the 142-keV $\gamma$ line coincident with 2.5$-$5.5 MeV protons is shown in the inset [of the figure], from which the half-life of the new nuclide $^{125}$Nd was extracted to be 0.60$\pm$0.15~s.'' This is the only reported half-life measurement, though a more recent paper by the same group reported the half-life as 0.65(15)~s \cite{2005Xu01}.

\subsection{$^{127}$Nd}\vspace{0.0cm}

Nitschke et al. discovered $^{127}$Nd in the 1983 paper ``New Beta-Delayed Proton Emitter in the Lanthanide Region'' \cite{1983Nit01}. A $^{40}$Ca beam from the Berkeley SuperHILAC bombarded a gas-cooled $^{92}$Mo target. $^{127}$Nd was produced in the fusion-evaporation reaction $^{92}$Mo($^{40}$Ca,$\alpha$n) and identified with the on-line isotope separator OASIS. ``From the growth and decay data, a half-life of 1.9~$\pm$~0.4~s was deduced, which is in good agreement with the calculated value of 1.8~s for $^{127}$Nd.'' This half-life is included in the currently accepted average value of 1.8(4)~s.

\subsection{$^{128}$Nd}\vspace{0.0cm}

In 1985 $^{128}$Nd was identified by Lister et al. in ``Deformation of Very Light Rare-Earth Nuclei'' \cite{1985Lis01}. A $^{40}$Ca beam from the Daresbury Laboratory Van de Graaff accelerator was incident on a $^{92}$Mo target and $^{128}$Nd was produced in the fusion-evaporation reaction $^{92}$Mo($^{40}$Ca,2p2n). Gamma rays, neutrons and charged particles were detected and new ground-state bands observed. ``This letter reports results on the ground-state bands in the even-even nuclei $^{128}_{58}$Ce$_{68}$, $^{132}_{60}$Nd$_{68,70,72}$, $^{134,136}_{62}$Sm$_{72,74}$, and $^{138,140}_{64}$Gd$_{74,76}$.'' An earlier identification and half-life measurement of 4(2)~s \cite{1983Nit01} was later refuted \cite{1985Wil01}.

\subsection{$^{129-133}$Nd}\vspace{0.0cm}

The first identification of $^{129}$Nd, $^{130}$Nd, $^{131}$Nd, $^{132}$Nd, and $^{133}$Nd, was reported in 1977 by Bogdanov et al. in ``New Neutron-Deficient Isotopes of Barium and Rare-Earth Elements'' \cite{1977Bog01}. The Dubna U-300 Heavy Ion Cyclotron accelerated a $^{32}$S beam which bombarded enriched targets of $^{102}$Pd and $^{106}$Cd. The isotopes were identified with the BEMS-2 isotope separator. ``In the present paper, isotopes were mainly identified by measuring the $\gamma$-ray and X-ray spectra of the daughter nuclei formed as a result of measuring the $\beta^+$ decay. In addition, the decay curves of the total $\beta$-activity of given isobars have been measured.'' The reported half-life of 5.9(6)~s for $^{129}$Nd is close to the accepted value of 4.9(2)~s. The half-life of 28(3)~s for $^{130}$Nd is within a factor of two of the currently accepted value of 13(3)~s. The half-lives of  24(3)~s ($^{131}$Nd) and 105(10)~s ($^{132}$Nd) are included in the currently accepted average values of 25.4(9)~s and 94(8)~s. The half-life of 70(10)~s for $^{133}$Nd corresponds to the presently adopted half-life.

\subsection{$^{134,135}$Nd}\vspace{0.0cm}

$^{134}$Nd and $^{135}$Nd were discovered by Abdurazakov et al. in the 1970 paper ``New Isotopes $^{133}$Pr, $^{134}$Nd, and $^{135}$Nd; Decay Schemes of $^{134}$Pr and $^{135}$Pr'' \cite{1970Abd01}. Spallation reactions were induced by 660 MeV protons irradiating a gadolinium target at the Joint Institute for Nuclear Reactions in Dubna, Russia. The reaction products were then chemically separated and identified by the measured $\gamma$-ray spectra. ``$^{134}$Nd was identified and its half-life determined via the strongest $\gamma$ rays of $^{134}$Pr (T = 17$\pm$2~min) with energies 409.2 and 639.0 keV.[The figure] shows the build-up and decline of the intensity of the 409.2~keV $\gamma$ lray of $^{134}$Pr. The half-life of was found from the difference of the curves (k1-k2) to be 8.5$\pm$1~min. A similar determination of this half-life using the 639.0 keV line gave 9$\pm$2 min.'' The currently accepted half-life of 8.5(15)~m is based on these two values. ``The half-life of $^{135}$Nd was found from the difference of the curves (k1-k2) to be 5.5 $\pm$ 0.5 min...'' This half-life has been assigned to an isomeric state but has not yet been confirmed.

\subsection{$^{136}$Nd}\vspace{0.0cm}

The assignment of $^{136}$Nd was reported in ``A New Neodymium Isotope (A = 136) and its Decay Properties'' by Zhelev et al. \cite{1968Zhe01}. A 660 MeV proton beam bombarded a gadolinium target and $^{136}$Nd was identified with $\gamma$- and $\beta$-ray spectra following chemical separation. ``We have previously reported the discovery of $^{137}$Nd, with a half life of 55.0 $\pm$ 1.5 min. Here we show that the $\beta^+$ and $\gamma$ radiations observed in the 55-min neodymium activity are not due to decay of $^{137}$Nd but to decay of a new isotope, $^{136}$Nd, and its daughter $^{136}$Pr.'' This half-life agrees with the currently accepted value of 50.65(33)~m. The paper mentioned in the quote is reference \cite{1965Gro01}. A 1-hour activity had previously been measured in neodymium \cite{1934Fer01,1935Ama01,1935Hev01} but no mass assignment were made.

\subsection{$^{137}$Nd}\vspace{0.0cm}

In 1970 $^{137}$Nd was reported in ``$^{135m}$Ce and $^{137m}$Nd: Isomeric States in N = 77 Isotones'' by Droste et al. \cite{1970Dro01}. The Dubna U-300 cyclotron accelerated $^{18}$O and $^{20,22}$Ne beams which bombarded tellurium and tin targets, respectively. $^{137}$Nd was produced in the fusion-evaporation reactions $^{119}$Sn($^{22}$Ne,4n), $^{122}$Sn($^{20}$Ne,5n), $^{122}$Sn($^{22}$Ne,7n), and $^{126}$Te($^{18}$O,7n). Conversion electron and $\gamma$-ray spectrometry was used to identify the isotope. ``In all cases, $\gamma$-rays of 108, 177, 233 and 285 keV which all decay with the same half-life, T$_{1/2}$ = 1.60$\pm$0.15~s were observed... That the new activity belongs indeed to the nucleide $^{137}$Nd is proved by (i) the energy difference of the K and L electron lines of the isomeric transition E${KL}$ = 37.0$\pm$0.8~keV as compared to the value 36.8 keV expected for Z = 60(Nd), and (ii) the shape of the excitation functions from which the number of evaporated neutrons can be determined'' The reported half-life corresponds to an isomeric state and is currently the accepted value for this state. Previously the 55.0(15)~m half-life of $^{136}$Nd had incorrectly been assigned to $^{137}$Nd \cite{1965Gro01}. A half-life of 35(5)~m corresponding to the ground state was reported already in 1935 \cite{1935McL01}, but no mass assignment was made.

\subsection{$^{138}$Nd}\vspace{0.0cm}

In 1964 Gromov et al. discovered $^{138}$Nd in ``Decay of Neutron-Deficient Neodymium Isotopes. A New Isotope Nd$^{138}$'' \cite{1964Gro01}. Erbium oxide or tantalum targets were irradiated with 660 MeV protons from the Dubna synchrocyclotron. Conversion electron and $\gamma$-ray spectra were measured to identify $^{138}$Nd. ``Observation of the E0 transition conversion lines, whose intensity decays with a $\approx$5-hour half-life, is experimental proof of the assumed isotope Nd$^{138}$ with this half-life.'' This half-life of about 5~h agrees with the currently accepted value of 5.04(9)~h. An activity of 22(2)~m, most likely the half-life of $^{139}$Nd, had previously been assigned incorrectly to $^{138}$Nd \cite{1951Sto01}.

\subsection{$^{139}$Nd}\vspace{0.0cm}

Stover reported the observation of $^{139}$Nd in the 1950 paper, ``New Neutron-Deficient Radioactive Isotopes of the Light Rare-Earth Region'' \cite{1951Sto01}. Praseodymium oxide was bombarded with 40 and 50 MeV protons from the Berkeley 184-in. cyclotron. Activities were measured with end-on type Geiger-M\"uller counters following chemical separation. ``Protons of energies 40 and 50 Mev gave a 22-min. activity, a 5.50-hr one, and the 3.3-day Nd$^{140}$... The 5.50-hr Nd was shown to be the grandparent of the 140-day Ce$^{139}$ and thus is Nd$^{139}$.'' The reported half-life of 5.5(2)~h is the currently accepted value for a long-lived isomeric state. The 22(2)~m activity is close to the half-life of the ground state (29.7(5)~m), however, Stover tentatively assigned this half-life to $^{138}$Nd.

\subsection{$^{140,141}$Nd}\vspace{0.0cm}

The discovery of $^{140}$Nd and $^{141}$Nd was reported by Wilkinson and Hicks in 1949: ``Radioactive Isotopes of the Rare Earth Elements II. Neodymium Isotopes'' \cite{1949Wil01}. Deuterons of 9 and 19~MeV and protons of 10~MeV from the Berkeley 60-in. cyclotron bombarded praseodymium targets. Positrons, X- and $\gamma$-rays were measured following chemical separation. ``The observed radiations of the 3.3-day activity are consistent with the isotope, Nd$^{140}$, decaying by orbital electron capture, in equilibrium with its positron emitting Pr$^{140}$ daughter.'' The reported half-life of 3.3(1)~d for $^{140}$Nd is consistent with the currently accepted value of 3.37(2)~d. ``145-Minute Nd$^{141}$:... Further, spectroscopic analysis showed that the 19.3-hour activity followed the praseodymium, while in the first active sample where the 145-minute decay was observed, praseodymium was below the limits of detection. The chemical identity of the 145-minute activity as neodymium is, therefore, fairly certain.'' The reported half-life of 145(3)~m is included in the currently accepted average value of 2.49(3)~h. Previous measurements of a 2.3~h \cite{1941Law01} and a 2.5~h half-life \cite{1942Kur01} were only published as conference abstracts.

\subsection{$^{142}$Nd}\vspace{0.0cm}

In 1924 Aston identified $^{142}$Nd in ``The Mass Spectra of Zirconium and Some Other Elements'' \cite{1924Ast02}. The Cavendish laboratory mass spectrograph was used to separate accelerated anode rays and identify their mass. ``Further experiments with neodymium (at. wt. 144.27) establish its principle isotopes as 142, 144, 146, with a possible (145).''

\subsection{$^{143}$Nd}\vspace{0.0cm}

The 1933 publication ``Constitution of Neodymium, Samarium, Europium, Gadolinium and Terbium'' by Aston reported the first observation of $^{143}$Nd \cite{1933Ast01}. Accelerated anode rays from the Cavendish laboratory mass spectrograph were analyzed and their masses identified. ``Three isotopes, 142, 14, 146, of neodymium (60) had already been identified by the first mass spectrograph. These are now found to be definitely in descending order of abundance and 143 and 145 also shown to be present.''

\subsection{$^{144}$Nd}\vspace{0.0cm}

In 1924 Aston identified $^{144}$Nd in ``The Mass Spectra of Zirconium and Some Other Elements'' \cite{1924Ast02}. The Cavendish laboratory mass spectrograph was used to separate accelerated anode rays and identify their mass. ``Further experiments with neodymium (at. wt. 144.27) establish its principle isotopes as 142, 144, 146, with a possible (145).''

\subsection{$^{145}$Nd}\vspace{0.0cm}

The 1933 publication ``Constitution of Neodymium, Samarium, Europium, Gadolinium and Terbium'' by Aston reported the first observation of $^{145}$Nd \cite{1933Ast01}. Accelerated anode rays from the Cavendish laboratory mass spectrograph were analyzed and their masses identified. ``Three isotopes, 142, 14, 146, of neodymium (60) had already been identified by the first mass spectrograph. These are now found to be definitely in descending order of abundance and 143 and 145 also shown to be present.'' A possible presence of $^{145}$Nd had been mentioned by Aston earlier \cite{1924Ast02,1925Ast02}.

\subsection{$^{146}$Nd}\vspace{0.0cm}

In 1924 Aston identified $^{146}$Nd in ``The Mass Spectra of Zirconium and Some Other Elements'' \cite{1924Ast02}. The Cavendish laboratory mass spectrograph was used to separate accelerated anode rays and identify their mass. ``Further experiments with neodymium (at. wt. 144.27) establish its principle isotopes as 142, 144, 146, with a possible (145).''

\subsection{$^{147}$Nd}\vspace{0.0cm}

In ``The Chemical Identification of Radioisotopes of Neodymium and of Element 61'', Marinsky et al. reported the discovery of $^{147}$Nd in 1947 \cite{1947Mar01}. Fission fragments from uranium fission were analyzed as part of the Manhattan Project. $^{147}$Nd was separated and identified with a synthetic organic cation exchanger of the sulfonated phenol-formaldehyde type. ``Radiochemical studies and fractionations have established the decay characteristics, genetic relation, and mass assignments of 11~d Nd$^{147}$, 3.7~y 61$^{147}$, 1.7~h Nd$^{(149)}$, and 47~h 61$^{149}$.'' The 11~d half-life is consistent with the currently accepted value of 10.98(1)~d. A 84~h half-life \cite{1938Poo02} reported previously was evidently incorrect. Other half-life measurements were only reported in conference abstracts \cite{1941Law01,1942Kur01}

\subsection{$^{148}$Nd}\vspace{0.0cm}

$^{148}$Nd was first identified by Dempster in 1937 as reported in ``Isotopic Constitution of Neodymium'' \cite{1937Dem02}. $^{148}$Nd was identified in the mass spectrum produced by a spark between neodymium electrodes. ``I have recently analyzed the ions from a spark between fairly pure neodymium electrodes, and find that the masses at 148 and 150 belong to this element.'' Dempster had previously observed lines at 148 and 150 \cite{1935Dem04} but was unable to assign the element. Aston subsequently had suggested that these line belong to neodymium \cite{1936Ast01}.

\subsection{$^{149}$Nd}\vspace{0.0cm}

The first detection of $^{149}$Nd was reported in 1938 by Pool and Quill in ``Radioactivity Induced in the Rare Earth Elements by Fast Neutrons'' \cite{1938Poo02}. Fast and slow neutrons were produced with 6.3 MeV deuterons from the University of Michigan cyclotron. Decay curves were measured with a Wulf string electrometer. ``In order to assign the 2.0-hr. and 84-hr. periods the relative abundance of the stable nuclei and the rate of formation of the radioactive nuclei (branching ratio) must be compared... Consequently, these data suggest, in view of no other guiding information, that the 84-hr. period should be attributed to Nd$^{147}$ and the 2-hr. period to Nd$^{149}$.'' This half-life agrees with the currently accepted value of 1.728(1)~h.

\subsection{$^{150}$Nd}\vspace{0.0cm}

$^{150}$Nd was first identified by Dempster in 1937 as reported in ``Isotopic Constitution of Neodymium'' \cite{1937Dem02}. $^{148}$Nd was identified in the mass spectrum produced by a spark between neodymium electrodes. ``I have recently analyzed the ions from a spark between fairly pure neodymium electrodes, and find that the masses at 148 and 150 belong to this element.'' Dempster had previously observed lines at 148 and 150 \cite{1935Dem04} but was unable to assign the element. Aston subsequently had suggested that these line belong to neodymium \cite{1936Ast01}.

\subsection{$^{151}$Nd}\vspace{0.0cm}

The first detection of $^{151}$Nd was reported in 1938 by Pool and Quill in ``Radioactivity Induced in the Rare Earth Elements by Fast Neutrons'' \cite{1938Poo02}. Fast and slow neutrons were produced with 6.3 MeV deuterons from the University of Michigan cyclotron. Decay curves were measured with a Wulf string electrometer. ``The very much greater neutron equivalent of the cyclotron is easily evident from the fact that in four hours of slow neutron bombardment three periods were easily evident, 21 min., 2.0 hr. and 84 hr... Since the 21-min. period is produced by slow but not by fast neutron bombardment and since Nd$^{150}$ is the heaviest neodymium isotope, it is reasonable to assign this activity to Nd$^{151}$.'' Although this half-life is somewhat larger than the presently accepted value (12.44(7)~m) it is still within a factor of two.

\subsection{$^{152}$Nd}\vspace{0.0cm}

$^{152}$Nd was first described in ``Rapid Isolation of Individual Rare Earths from Fission and Identification of $^{152}$Nd'' by Wakat and Griffin in 1969 \cite{1969Wak01}. The isotope was produced by neutron induced fission of $^{235}$U. Decay curves were measured with a 4$\pi\beta$ detector after element separation in a resin column. ``The decay curves can be resolved into 2 components: a growth and decay component which is consistent with the activity of the daughter in a 11.3$\rightarrow$4.1-min genetic pair, and a 52-min activity... These data confirm the presence of 11.3 $\pm$ 0.4-min $^{152}$Nd, even though characteristic radiations of $^{152}$Nd have not been observed directly.'' This half-life is included in the currently accepted average value of 11.2(2)~m. A previous result had been reported in a conference abstract \cite{1969Hof01} and only a month later an independent identification of $^{152}$Nd was submitted \cite{1970Cha01}.

\subsection{$^{153}$Nd}\vspace{0.0cm}

In 1987, Greenwood et al. identified $^{153}$Nd in the paper entitled ``Identification of New Neutron-Rich Rare-Earth Isotopes Produced in $^{252}$Cf Fission'' \cite{1987Gre01}. Spontaneous fission fragments from a $^{252}$Cf source were measured with the isotope separation on line (ISOL) system at the Idaho National Engineering Laboratory. $^{153}$Nd was identified by mass separation and the measurement of K x-rays. ``Identification of the $^{153}$Nd isotope was first reported by Pinston et al. In the present work, however, a total of 48 $\gamma$-ray transitions could be associated with this decay, compared with the eight transitions reported earlier. The half-life value was obtained from an average of individual values involving the Pm K x rays and the 32.2-, 105.5-, 345.0-, 418.3-, and 976.1-keV $\gamma$ rays.'' The reported half-life of 28.9(4)~s is included in the currently accepted average value of 31.6(10)~s. The quoted Pinston reference was only published in a conference proceedings nine years earlier \cite{1978Pin01}.

\subsection{$^{154}$Nd}\vspace{0.0cm}

$^{154}$Nd was discovered by Wilhelmy et al. in 1970 in ``Ground-State Bands in Neutron-Rich Even Te, Xe, Ba, Ce, Nd, and Sm Isotopes Produced in the Fission of $^{252}$Cf'' \cite{1970Wil01}. The isotope was observed in the spontaneous fission of a $^{252}$Cf source. The identification was based on coincidence measurements of both fission fragments and K-x-rays. Gamma-rays for several isotopes were measured and the results were only displayed in a table. The first 4 transitions up to the decay of the 8$^+$ state were measured for $^{154}$Nd. A month earlier another measurement of spontaneous fission fragments extracted isomeric states using the four-parameter method. However, this method does not have a unique mass identification and the atomic number is only deduced from the most probable Z, and is thus not considered sufficient evidence for a discovery.

\subsection{$^{155}$Nd}\vspace{0.0cm}

Okano et al. identified $^{155}$Nd in the 1986 paper ``The Half-Life of a New Isotope, $^{155}$Nd'' \cite{1986Oka01}. An enriched $^{235}$U target was irradiated with neutrons from the Kyoto University Reactor and $^{155}$Nd was identified with a helium-jet system and the on-line isotope separator KUR-ISOL. ``From the analyses of the spectra mentioned, four $\gamma$-rays of 67.5$\pm$0.3, 180.7$\pm$0.2, 418.9$\pm$0.3, and 955.1$\pm$0.3 keV and two X-rays of 38.4$\pm$0.4 (Pm K$_\alpha$) and 44.2$\pm$1.0 keV (Pm K$_\beta$) were assigned to the decay of $^{155}$Nd.'' The reported half-life of 9.5(7)~s agrees with the currently accepted value of 8.9(2)~s. About six months later Greenwood et al. claimed the first observation of $^{155}$Nd \cite{1987Gre01} apparently not aware of the Okano et al. publication.

\subsection{$^{156}$Nd}\vspace{0.0cm}

In 1987, Greenwood et al. identified $^{156}$Nd in the paper entitled ``Identification of New Neutron-Rich Rare-Earth Isotopes Produced in $^{252}$Cf Fission'' \cite{1987Gre01}. Spontaneous fission fragments from a $^{252}$Cf source were measured with the isotope separation on line (ISOL) system at the Idaho National Engineering Laboratory. $^{156}$Nd was identified by mass separation and the measurement of K x-rays. ``The half-life value for $^{156}$Nd was obtained as an average of individual measurements involving the Pm K x rays and the 84.8- and 150.7-keV $\gamma$ rays which we can definitely associate with this decay at this time.'' The reported half-life of 5.47(11)~s is included in the currently accepted average value of 5.49(7)~s.

\section{Summary}
The discoveries of the known zinc, selenium, bromine, and neodymium isotopes have been compiled and the methods of their production discussed.

While the discovery of the neutron-rich and neutron-deficient zinc isotopes was uncontroversial, the identification of the stable and near-stable isotopes was more difficult. The stable isotopes $^{64}$Zn, $^{66}$Zn, $^{68}$Zn, and $^{70}$Zn were initially misidentified as $^{63}$Zn, $^{65}$Zn, $^{67}$Zn, $^{69}$Zn, respectively. Then $^{65}$Zn and $^{69}$Zn as well as $^{63}$Zn were incorrectly reported to be stable. Before $^{65}$Zn and $^{69}$Zn were finally identified correctly an erroneous half-life was reported for $^{65}$Zn and the half-life of $^{69}$Zn was at first assigned to $^{65}$Zn.

The identification of the selenium isotopes was difficult. The half-lives of $^{68}$Se, $^{69}$Se and $^{86}$Se were initially incorrect and the half-lives of $^{79}$Se and $^{81}$Se could at first not be assigned to a unique mass. The half-lives of $^{70}$Se and $^{86}$Se were first assigned to $^{71}$Se and $^{87}$Se, respectively. Finally, the first studies of $^{75}$Se and $^{84}$Se were reported in classified reports of the Manhattan Project.

The limit for observing long lived bromine isotopes beyond the proton dripline which can be measured by implantation decay studies has most likely been reached with the discovery of $^{70}$Br and the observation that of $^{69}$Br has an upper half-life limit of 100~ns. The identification of several of the bromine isotopes was difficult and many half-lives were initially determined without a mass assignment ($^{73}$Br, $^{80}$Br, $^{84}$Br, $^{85}$Br, $^{87}$Br, and $^{89}$Br). The half-life of $^{83}$Br was at first incorrectly assigned to $^{78}$Br.

The half-lives of two neodymium isotopes ($^{128}$Nd and $^{147}$Nd) were initially measured incorrectly and the half-lives of $^{136}$Nd and $^{137}$Nd were first measured without a unique mass assignment. Also, the half-life of $^{139}$Nd was at first assigned to $^{138}$Nd. Finally, the discoveries of $^{141}$Nd, $^{147}$Nd, and $^{153}$Nd were published in refereed journals only several years after the first announcements in conference proceedings.

\ack

The main research on the individual elements were performed by JLG (zinc and neodymium), JC (selenium) and JK (bromium). This work was supported by the National Science Foundation under grant No. PHY06-06007 (NSCL).

\bibliography{../isotope-discovery-references}

\begin{thebibliography}{100}
\expandafter\ifx\csname url\endcsname\relax
  \def\url#1{\texttt{#1}}\fi
\expandafter\ifx\csname urlprefix\endcsname\relax\def\urlprefix{URL }\fi

\bibitem{2009Gin01}
G.~Q. Ginepro, J.~Snyder, M.~Thoennessen, At. Data. Nucl. Data. Tables 95
  (2009) 805.

\bibitem{2003Aud01}
G.~Audi, O.~Bersillon, J.~Blachot, A.~H. Wapstra, Nucl. Phys. A 729 (2003) 3.

\bibitem{2008ENS01}
{ ENSDF, Evaluated Nuclear Structure Data File, maintained by the National
  Nuclear Data Center at Brookhaven National Laboratory, published in Nuclear
  Data Sheets (Academic Press, Elsevier Science)}.

\bibitem{1970Joh01}
W.~John, F.~W. Guy, J.~J. Wesolowski, Phys. Rev. C 2 (1970) 1451.

\bibitem{1946TPP01}
{The Plutonium Project, J. Am. Chem. Soc. 68 (1946) 2411; Rev. Mod. Phys. 18
  (1946) 513}.

\bibitem{1951Cor01}
{ C. D. Coryell and N. Sugarman (Editors), Radiochemical Studies: The Fission
  Products, National Nuclear Energy Series IV, 9, (McGraw-Hill, New York 1951)
  }.

\bibitem{2008NSR01}
{ http://www.nndc.bnl.gov/nsr/ NSR, Nuclear Science References, maintained by
  the National Nuclear Data Center at Brookhaven National Laboratory }.

\bibitem{1940Liv01}
J.~J. Livingood, G.~T. Seaborg, Rev. Mod. Phys. 12 (1940) 30.

\bibitem{1944Sea01}
G.~T. Seaborg, Rev. Mod. Phys. 16 (1944) 1.

\bibitem{1948Sea01}
G.~Seaborg, I.~Perlman, Rev. Mod. Phys. 20 (1948) 585.

\bibitem{1953Hol02}
J.~M. Hollander, I.~Perlman, G.~T. Seaborg, Rev. Mod. Phys. 25 (1953) 469.

\bibitem{1958Str01}
D.~Strominger, J.~M. Hollander, G.~T. Seaborg, Rev. Mod. Phys. 30 (1958) 585.

\bibitem{1967Led01}
{ C. M. Lederer, J. M. Hollander, I. Perlman, Table of Isotopes, 6$^{th}$
  Edition, John Wiley \& Sons 1967}.

\bibitem{1942Ast01}
{ F. W. Aston, Mass Spectra and Isotopes, 2$^{nd}$ Edition, Longmans, Green \&
  Co., New York 1942 }.

\bibitem{2007Gor01}
S.~Goriely, M.~Samyn, J.~M. Pearson, Phys. Rev. C 75 (2007) 064312.

\bibitem{2004Tho01}
M.~Thoennessen, Rep. Prog. Phys. 67 (2004) 1187.

\bibitem{2005Bla01}
B.~Blank, A.~Bey, G.~Canchel, C.~Dossat, A.~Fleury, J.~Giovinazzo, I.~Matea,
  N.~Adimi, F.~{De Oliveira}, I.~Stefan, G.~Georgiev, S.~Gr\'evy, J.~C. Thomas,
  C.~Borcea, D.~Cortina, M.~Caamano, M.~Stanoiu, F.~Aksouh, B.~A. Brown, F.~C.
  Barker, W.~A. Richter, Phys. Rev. Lett. 94 (2005) 232501.

\bibitem{2001Gio01}
J.~Giovinazzo, B.~Blank, C.~Borcea, M.~Chartier, S.~Czajkowski, G.~{de France},
  R.~Grzywacz, Z.~Janas, M.~Lewitowicz, F.~{de Oliveira Santos}, M.~Pf\"utzner,
  M.~S. Pravikoff, J.~C. Thomas, Eur. Phys. J. A 11 (2001) 247.

\bibitem{1976Vie01}
D.~J. Vieira, D.~F. Sherman, M.~S. Zisman, R.~A. Gough, J.~Cerny, Phys. Lett. B
  60 (1976) 261.

\bibitem{1986Set01}
K.~K. Seth, S.~Iversen, M.~Kaletka, D.~Barlow, A.~Saha, R.~Soundranayagam,
  Phys. Lett. B 173 (1986) 397.

\bibitem{1981Hon01}
J.~Honkanen, M.~Kortelahti, K.~Eskola, K.~Vierinen, Nucl. Phys. A 366 (1981)
  109.

\bibitem{1981Ara01}
Y.~Arai, M.~Fujioka, E.~Tanaka, T.~Shinozuka, H.~Miyatake, M.~Yohii,
  T.~Ishimatsu, Phys. Lett. B 104 (1981) 186.

\bibitem{1955Lin01}
L.~Lindner, G.~A. Brinkman, Physica 21 (1955) 747.

\bibitem{1955Cum01}
J.~B. Cumming, Bull. Am .Phys. Soc. 30 (2007) L2.

\bibitem{1959Cum01}
J.~B. Cumming, Phys. Rev. 114 (2008) 1600.

\bibitem{1948Mil01}
D.~R. Miller, R.~C. Thompson, B.~B. Cunningham, Phys. Rev. 74 (1948) 347.

\bibitem{1937Bot03}
W.~Bothe, W.~Gentner, Naturwiss. 25 (1937) 191.

\bibitem{1935Ste01}
G.~Stenvinkel, E.~Svensson, Nature 135 (1935) 955.

\bibitem{1922Dem01}
A.~J. Dempster, Phys. Rev. 20 (1922) 631.

\bibitem{1921Dem01}
A.~J. Dempster, Proc. Nat. Acad. Sci. 7 (1921) 45.

\bibitem{1939Liv03}
J.~J. Livingood, G.~T. Seaborg, Phys. Rev. 55 (1939) 457.

\bibitem{1938Per01}
C.~Perrier, M.~Santangelo, E.~Segre, Phys. Rev. 53 (1938) 104.

\bibitem{1936Hey01}
F.~A. Heyn, Nature 138 (1936) 723.

\bibitem{1938Sag01}
R.~Sagane, S.~Kojima, M.~Ikawa, Phys. Rev. 54 (1938) 149.

\bibitem{1928Ast01}
F.~W. Aston, Nature 122 (1928) 345.

\bibitem{1937Hey02}
F.~A. Heyn, Physica 4 (1937) 1224.

\bibitem{1936Liv01}
J.~J. Livingood, Phys. Rev. 50 (1936) 425.

\bibitem{1955LeB01}
J.~M. LeBlanc, J.~M. Cork, S.~B. Burson, Phys. Rev. 97 (1955) 750.

\bibitem{1951Sie01}
J.~M. Siegel, L.~E. Glendenin, Radiochemical Studies: The Fission Products,
  Paper 53, p. 549, National Nuclear Energy Series IV, 9, McGraw-Hill, New
  York, 1951.

\bibitem{1972Erd01}
B.~R. Erdal, L.~Westgaard, J.~Zylicz, E.~Roeckl, {the ISOLDE Collaboration},
  Nucl. Phys. A 194 (1972) 449.

\bibitem{1974Gra01}
B.~Grapengiesser, E.~Lund, G.~Rudstam, J. Inorg. Nucl. Chem. 36 (1974) 2409.

\bibitem{1977Ale01}
K.~Aleklett, E.~Lund, G.~Nyman, G.~Rudstam, Nucl. Phys. A 285 (1977) 1.

\bibitem{1975Ale01}
K.~Aleklett, G.~Nyman, G.~Rudstam, Nucl. Phys. A 246 (1975) 425.

\bibitem{1981Rud01}
G.~Rudstam, P.~Aagaard, P.~Hoff, B.~Johansson, H.-U. Zwicky, Nucl. Instrum.
  Meth. 186 (1981) 365.

\bibitem{1976Rud01}
G.~Rudstam, E.~Lund, Phys. Rev. C 13 (1976) 321.

\bibitem{1977Rud01}
G.~Rudstam, E.~Lund, Nucl. Sci. Eng. 64 (1977) 749.

\bibitem{1991Kra01}
K.-L. Kratz, H.~Gabelmann, P.~Moller, B.~Pfeiffer, H.~L. Ravn, A.~Wohr, the
  ISOLDE~Collaboration, Z. Phys. A 340 (1991) 419.

\bibitem{1997Ber01}
M.~Bernas, C.~Engelmann, P.~Armbruster, S.~Czajkowski, F.~Ameil,
  C.~B\"ockstiegel, P.~Dessagne, C.~Donzaud, H.~Geissel, A.~Heinz, Z.~Janas,
  C.~Kozhuharov, C.~Mieh\'e, G.~M\"unzenberg, M.~Pf\"utzner, W.~Schwab,
  C.~St\'ephan, K.~S\"ummerer, L.~Tassan-Got, B.~Voss, Phys. Lett. B 415 (1997)
  111.

\bibitem{2010Ohn01}
T.~Ohnishi, T.~Kubo, K.~Kusaka, A.~Yoshida, K.~Yoshida, M.~Ohtake, N.~Fukuda,
  H.~Takeda, D.~Kameda, K.~Tanaka, N.~Inabe, Y.~Yanagisawa, Y.~Gono,
  H.~Watanabe, H.~Otsu, H.~Baba, T.~Ichihara, Y.~Yamaguchi, M.~Takechi,
  S.~Nishimura, H.~Ueno, A.~Yoshimi, H.~Sakurai, T.~Motobayashi, T.~Nakao,
  Y.~Mizoi, M.~Matsushita, K.~Ieki, N.~Kobayashi, K.~Tanaka, Y.~Kawada,
  N.~Tanaka, S.~Deguchi, Y.~Satou, Y.~Kondo, T.~Nakamura, K.~Yoshinaga,
  C.~Ishii, H.~Yoshii, Y.~Miyashita, N.~Uematsu, Y.~Shiraki, T.~Sumikama,
  J.~Chiba, E.~Ideguchi, A.~Saito, T.~Yamaguchi, I.~Hachiuma, T.~Suzuki,
  T.~Moriguchi, A.~Ozawa, T.~Ohtsubo, M.~A. Famiano, H.~Geissel, A.~S.
  Nettleton, O.~B. Tarasov, D.~Bazin, B.~M. Sherrill, S.~L. Manikonda, J.~A.
  Nolen, J. Phys. Soc. Japan 79 (2010) 073201.

\bibitem{1977Gru01}
H.~A. Grunder, F.~Selph, Annu. Rev. Nucl. Sci. 27 (1977) 353.

\bibitem{2005Sto01}
A.~Stolz, T.~Baumann, N.~H. Frank, T.~N. Ginter, G.~W. Hitt, E.~Kwan, M.~Mocko,
  W.~Peters, A.~Schiller, C.~S. Sumithrarachchi, M.~Thoennessen, Phys. Lett. B
  627 (2005) 32.

\bibitem{1993Bat01}
J.~C. Batchelder, D.~M. Moltz, T.~J. Ognibene, M.~W. Rowe, J.~Cerny, Phys. Rev.
  C 47 (1993) 2038.

\bibitem{1993Win01}
J.~A. Winger, D.~P. Bazin, W.~Benenson, G.~M. Crawley, D.~J. Morrissey, N.~A.
  Orr, R.~Pfaff, B.~M. Sherrill, M.~Thoennessen, S.~J. Yennello, B.~M. Young,
  Phys. Rev. C 48 (1993) 3097.

\bibitem{1991Moh01}
M.~F. Mohar, D.~Bazin, W.~Benenson, D.~J. Morrissey, N.~A. Orr, B.~M. Sherrill,
  D.~Swan, J.~A. Winger, A.~C. Mueller, D.~Guillemaud-Mueller, Phys. Rev. Lett.
  66 (1991) 1571.

\bibitem{1989Lan01}
{T. F. Lang, J. D. Robertson, D. M. Moltz, J. E. Reiff, J. Batchelder, J.
  Cerny, C. Beausang, M. A. Deleplanque, R. M. Diamond, F. S. Stephens, Bull.
  Am. Phys. Soc. 34, No. 8, 1801, AC9 (1989)}.

\bibitem{1990Lis01}
C.~J. Lister, P.~J. Ennis, A.~A. Chishti, B.~J. Varley, W.~Gelletly, H.~G.
  Price, A.~N. James, Phys. Rev. C 42 (1990) R1191.

\bibitem{1972Bil01}
A.~N. Bilge, G.~G.~J. Boswell, J. Inorg. Nucl. Chem. 34 (1972) 407.

\bibitem{1976LaB01}
J.~J. {La Brecque}, I.~L. Preiss, J. Inorg. Nucl. Chem. 38 (1976) 2139.

\bibitem{1974Nol01}
E.~Nolte, Y.~Shida, W.~Kutschera, R.~Prestele, H.~Morinaga, Z. Phys. 268 (1974)
  267.

\bibitem{1973Pre01}
I.~L. Preiss, J.~J. {La Brecque}, H.~Bakhru, R.~I. Morse, Nucl. Phys. A 205
  (1973) 619.

\bibitem{1950Hop01}
H.~H. {Hopkins Jr.}, Phys. Rev. 77 (1950) 717.

\bibitem{1948Hop01}
H.~H. {Hopkins Jr.}, B.~B. Cunningham, Phys. Rev. 73 (1948) 1406.

\bibitem{1957Bey01}
J.~Beydon, H.~Faraggi, I.~Gratot, M.~LePape, Compt. Rend. Acad. Sci. 244 (1957)
  586.

\bibitem{1948Cow01}
W.~S. Cowart, M.~L. Pool, D.~A. McCown, L.~L. Woodward, Phys. Rev. 73 (1948)
  1454.

\bibitem{1922Ast03}
F.~W. Aston, Nature 110 (1922) 664.

\bibitem{1947Fri01}
H.~N. Friedlander, L.~Seren, S.~H. Turkel, Phys. Rev. 73 (1947) 23.

\bibitem{1946ARI01}
{ Availability of Radioactive Isotopes, Announcement from Headquarters,
  Manhattan Project, Washing, D.C., Science 103 (1946) 697}.

\bibitem{1944Ges01}
{ H. Gest and L. E. Glendenin, Plutonium Project Report CC-3389, p.11 (1946)}.

\bibitem{1950Fla01}
A.~Flammersfeld, W.~Herr, Z. Naturforsch 5a (1950) 569.

\bibitem{1950Fla02}
A.~Flammersfeld, C.~Ythier, Z. Naturforsch 5a (1950) 401.

\bibitem{1952Har01}
W.~A. Hardy, G.~Silvey, C.~H. Townes, Phys. Rev. 85 (1952) 494.

\bibitem{1948Waf01}
H.~Waffler, O.~Hirzel, Helv. Phys. Acta 21 (1948) 200.

\bibitem{1937Sne01}
A.~H. Snell, Phys. Rev. 52 (1937) 1007.

\bibitem{1940Lan01}
A.~{Langsdorf Jr.}, E.~Segre, Phys. Rev. 57 (1940) 105.

\bibitem{1939Bot01}
W.~Bothe, W.~Gentner, Z. Phys. 112 (1939) 45.

\bibitem{1960Sat01}
J.~E. Sattizahn, J.~D. Knight, J. Inorg. Nucl. Chem. 12 (1960) 206.

\bibitem{1946Gle01}
{ L. E. Glendenin, NNES-PPR, Vol. 9B, Paper No. 7.3.1 (1946) }.

\bibitem{1944Edw01}
{ R. Edwards and H. Gest and T. Davies, Plutonium Project Report CC-2485, p. 3
  (Dec. 1944)}.

\bibitem{1951Edw01}
R.~R. Edwards, H.~Gest, T.~H. Davies, Radiochemical Studies: The Fission
  Products, p. 237, National Nuclear Energy Series IV, 9, McGraw-Hill, New
  York, 1951.

\bibitem{1973Tam01}
T.~Tamai, R.~Matsushita, J.~Takada, Y.~Kiso, J. Inorg. Nucl. Chem. Lett. 9
  (1973) 1145.

\bibitem{1969Her01}
G.~Herrmann, H.~O. Denschlag, Annu. Rev. Nucl. Sci. 19 (1969) 1.

\bibitem{1968Tom01}
L.~Tomlinson, M.~H. Hurdus, J. Inorg. Nucl. Chem. 30 (1968) 1995.

\bibitem{1970Mar01}
P.~del Marmol, D.~C. Perricos, J. Inorg. Nucl. Chem. 32 (1970) 705.

\bibitem{1971Tom01}
L.~Tomlinson, M.~H. Hurdus, J. Inorg. Nucl. Chem. 33 (1971) 3609.

\bibitem{1994Ber01}
M.~Bernas, S.~Czajkowski, P.~Armbruster, H.~Geissel, P.~Dessagne, C.~Donzaud,
  H.-R. Faust, E.~Hanelt, A.~Heinz, M.~Hesse, C.~Kozhuharov, C.~Mieh\'e,
  G.~M\"unzenberg, M.~Pf\"utzner, C.~R\"ohl, K.~H. Schmidt, W.~Schwab,
  C.~St\'ephan, K.~S\"ummerer, L.~Tassan-Got, B.~Voss, Phys. Lett. B 331 (1994)
  19.

\bibitem{1975Asg01}
M.~Asghar, J.~P. Gautheron, G.~Bailleul, J.~P. Bocquet, J.~Greif, H.~Schrader,
  G.~Siegert, C.~Ristori, J.~Crancon, G.~I. Crawford, Nucl. Phys. A 247 (1975)
  359.

\bibitem{1995Bla01}
B.~Blank, S.~Andriamonje, S.~Czajkowski, F.~Davi, R.~D. Moral, J.~P. Dufour,
  A.~Fleury, A.~Musquere, M.~S. Pravikoff, R.~Grzywacz, Z.~Janas,
  M.~Pf\"utzner, A.~Grewe, A.~Heinz, A.~Junghans, M.~Lewitowicz, J.~E.
  Sauvestre, C.~Donzaud, Phys. Rev. Lett. 74 (1995) 4611.

\bibitem{1978Alb01}
D.~E. Alburger, Phys. Rev. C 18 (1978) 1875.

\bibitem{1981Vos01}
B.~Vosicki, T.~Bj\"{o}rnstad, L.~C. Carraz, J.~Heinemeier, H.~L. Ravn, Nucl.
  Instrum. Meth. 186 (1981) 307.

\bibitem{1970Nol01}
E.~Nolte, W.~Kutschera, Y.~Shida, H.~Morinaga, Phys. Lett. B 33 (1970) 294.

\bibitem{1971Dor01}
T.~A. Doron, M.~Blann, Nucl. Phys. A 161 (1971) 12.

\bibitem{1970Mur01}
G.~Murray, W.~J.~K. White, J.~C. Willmott, R.~F. Entwistle, Nucl. Phys. A 142
  (1970) 21.

\bibitem{1953Hol01}
J.~M. Hollander, Phys. Rev. 92 (1953) 518.

\bibitem{1960But01}
F.~D.~S. Butement, G.~G.~J. Boswell, J. Inorg. Nucl. Chem. 16 (1960) 10.

\bibitem{1948Woo02}
L.~L. Woodward, D.~A. McCown, M.~L. Pool, Phys. Rev. 74 (1948) 870.

\bibitem{1952Ful01}
S.~C. Fultz, M.~L. Pool, Phys. Rev. 86 (1952) 347.

\bibitem{1937Bot01}
W.~Bothe, W.~Gentner, Naturwiss. 25 (1937) 126.

\bibitem{1937Cha01}
W.~Y. Chang, M.~Goldhaber, R.~Sagane, Nature 139 (1937) 962.

\bibitem{1937Hey01}
F.~A. Heyn, Nature 138 (1937) 842.

\bibitem{1935Kur01}
B.~Kurtchatov, L.~Kurtchatov, L.~Myssowky, L.~Roussinov, Compt. Rend. Acad.
  Sci. 200 (1935) 1201.

\bibitem{1920Ast02}
F.~W. Aston, Nature 105 (1920) 547.

\bibitem{1937Bot02}
W.~Bothe, W.~Gentner, Naturwiss. 25 (1937) 90.

\bibitem{1935Ama01}
E.~Amaldi, O.~D'Agostino, F.~Rasetti, E.~Segre, Proc. Roy. Soc. A 149 (1935)
  522.

\bibitem{1940Str01}
F.~Strassmann, O.~Hahn, Naturwiss. 28 (1940) 817.

\bibitem{1943Bor01}
H.~J. Born, W.~Seelmann-Eggebert, Naturwiss. 31 (1943) 86.

\bibitem{1939Hah01}
O.~Hahn, F.~Strassmann, Naturwiss. 27 (1939) 89.

\bibitem{1939Dod01}
R.~W. Dodson, R.~D. Fowler, Phys. Rev. 55 (1939) 880.

\bibitem{1962Ste01}
A.~F. Stehney, E.~P. Steinberg, Phys. Rev. 127 (1962) 563.

\bibitem{1949Sug01}
N.~Sugarman, J. Chem. Phys. 17 (1949) 11.

\bibitem{1959Per01}
G.~J. Perlow, A.~F. Stehney, Phys. Rev. 113 (1959) 1269.

\bibitem{1947Red01}
W.~C. Redman, D.~Saxon, Phys. Rev. 72 (1947) 570.

\bibitem{1948Hug01}
D.~J. Hughes, J.~Dabbs, A.~Cahn, D.~Hall, Phys. Rev. 73 (1948) 111.

\bibitem{1947Sug01}
N.~Sugarman, J. Chem. Phys. 15 (1947) 544.

\bibitem{1974Kra01}
K.~L. Kratz, G.~Herrmann, Nucl. Phys. A 229 (1974) 179.

\bibitem{2000Sou01}
G.~A. Souliotis, Physica Scripta T88 (2000) 153.

\bibitem{1997Ber02}
M.~Bernas, P.~Armbruster, S.~Czajkowski, C.~Donzaud, H.~Geissel, F.~Ameil,
  P.~Dessagne, C.~Engelmann, A.~Heinz, Z.~Janas, C.~Kozhuharov, C.~Miehe,
  G.~M\"unzenberg, M.~Pf\"utzner, C.~B\"ocksteigel, K.~H. Schmidt, W.~Schwab,
  C.~St\'ephan, K.~S\"ummerer, L.~Tassan-Got, B.~Voss, Nucl. Phys. A 616 (1997)
  352c.

\bibitem{1999Xu01}
S.~W. Xu, Z.~K. Li, Y.~X. Xie, Q.~Y. Pan, Y.~Yu, J.~Adam, C.~F. Wang, J.~P.
  Xing, Q.~Y. Hu, S.~H. Li, H.~Y.Chen, T.~M. Zhang, G.~M. Jin, Y.~X. Luo,
  Y.~Penionzhkevich, Y.~Gangrsky, Phys. Rev. C 60 (1999) 061302.

\bibitem{2005Xu01}
S.~W. Xu, Z.~K. Li, Y.~X. Xie, Q.~Y. Pan, W.~X. Huang, X.~D. Wang, Y.~Yu, Y.~B.
  Xing, Phys. Rev. C 71 (2005) 054318.

\bibitem{1983Nit01}
J.~M. Nitschke, M.~D. Cable, W.~D. Zeitz, Z. Phys. A 312 (1983) 265.

\bibitem{1985Lis01}
C.~J. Lister, B.~J. Varley, R.~Moscrop, W.~Gellently, P.~J. Nolan, D.~J.~G.
  Love, P.~J. Bishop, A.~Kirwan, D.~J. Thornley, L.~Ying, R.~Wadsworth, J.~M.
  O'Donell, H.~G. Price, A.~H. Nelson, Phys. Rev. Lett. 55 (1985) 810.

\bibitem{1985Wil01}
P.~A. Wilmarth, J.~M. Nitschke, P.~K. Lemmertz, R.~B. Firestone, Z. Phys. A 321
  (1985) 179.

\bibitem{1977Bog01}
D.~D. Bogdanov, A.~V. Demyanov, V.~A. Karnaukhov, L.~A. Petrov, A.~Plohocki,
  V.~G. Subbotin, J.~Voboril, Nucl. Phys. A 275 (1977) 229.

\bibitem{1970Abd01}
A.~A. Abdurazakov, R.~Arlt, R.~Babadzhanov, G.~Baier, V.~A. Morozov, G.~Musiol,
  K.~Tyrroff, H.~Strusny, Izv. Akad. Nauk SSSR, Ser. Fiz 34 (1970) 796.

\bibitem{1968Zhe01}
Z.~Zhelev, V.~G. Kallinnikov, J.~Liptak, L.~Peker, Izv. Akad. Nauk SSSR 32
  (1968) 1610.

\bibitem{1965Gro01}
K.~Gromov, V.~Kalinnikov, V.~Kuznetsov, N.~Lebedev, G.~Musiol, E.~Herrmann,
  Z.~Zhelev, B.~Dzhelepov, A.~Kudryavtseva, Nucl. Phys. 73 (1965) 65.

\bibitem{1934Fer01}
E.~Fermi, E.~Amaldi, O.~D'Agostino, F.~Rasetti, E.~Segre, Proc. Roy. Soc. A 146
  (1934) 483.

\bibitem{1935Hev01}
G.~Hevesy, H.~Levi, Nature 136 (1935) 103.

\bibitem{1970Dro01}
C.~Droste, W.~Neubert, J.~Lewitowicz, S.~Chojnacki, T.~Morek, Z.~Wilhelmi,
  K.~F. Alexander, Nucl. Phys. A 152 (1970) 579.

\bibitem{1935McL01}
J.~C. McLennan, L.~G. Grimmett, J.~Read, Nature 135 (1935) 147.

\bibitem{1964Gro01}
K.~Y. Gromov, A.~S. Danagulyan, L.~N. Kikityuk, V.~V. Murav'eva, A.~A. Sorokin,
  M.~Z. Shtal', V.~Shpinel', Zh. Eksperim. i Teor. Fiz. 47 (1964) 1644.

\bibitem{1951Sto01}
B.~J. Stover, Phys. Rev. 81 (1951) 8.

\bibitem{1949Wil01}
G.~Wilkinson, H.~G. Hicks, Phys. Rev. 75 (1949) 1687.

\bibitem{1941Law01}
H.~B. Law, M.~L. Pool, J.~D. Kurbatov, L.~Quill, Phys. Rev. 59 (1941) 936.

\bibitem{1942Kur01}
J.~D. Kurbatov, D.~C. MacDonald, M.~L. Pool, L.~Quill, Phys. Rev. 61 (1942)
  106.

\bibitem{1924Ast02}
F.~W. Aston, Nature 114 (1924) 273.

\bibitem{1933Ast01}
F.~W. Aston, Nature 132 (1933) 930.

\bibitem{1925Ast02}
F.~W. Aston, Phil. Mag. 49 (1925) 1191.

\bibitem{1947Mar01}
J.~A. Marinsky, L.~Glendenin, C.~Coryell, J. Am. Chem. Soc. 69 (1947) 2781.

\bibitem{1938Poo02}
M.~L. Pool, L.~L. Quill, Phys. Rev. 53 (1938) 437.

\bibitem{1937Dem02}
A.~J. Dempster, Phys. Rev. 51 (1937) 289.

\bibitem{1935Dem04}
A.~J. Dempster, Proc. Am. Phys. Soc. 75 (1935) 735.

\bibitem{1936Ast01}
F.~W. Aston, Nature 137 (1936) 613.

\bibitem{1969Wak01}
A.~Wakat, C.~Griffin, Radiochem. Radioanal. Lett. 2 (1969) 351.

\bibitem{1969Hof01}
D.~C. Hoffman, F.~O. Lawrence, W.~Daniels, Bull. Am. Phys. Soc. 14 (1969) 1225.

\bibitem{1970Cha01}
R.~Chapman, W.~McLatchie, J.~Kitching, Phys. Lett. B 31 (1970) 292.

\bibitem{1987Gre01}
R.~C. Greenwood, R.~A. Anderl, J.~D. Cole, H.~Willmes, Phys. Rev. C 35 (1987)
  1965.

\bibitem{1978Pin01}
J.~A. Pinston, F.~Schussler, E.~M. J.~P. Zirnheld, V.~Raut, G.~J. Costa,
  A.~Hanni, R.~Seltz, Proc. 6th Intern. Conf. on Atomic Masses and Fund. Const.
  (1978) 493.

\bibitem{1970Wil01}
J.~B. Wilhelmy, S.~G. Thompson, R.~C. Jared, E.~Cheifetz, Phys. Rev. Lett. 25
  (1970) 1122.

\bibitem{1986Oka01}
K.~Okano, Y.~Kawase, K.~Aoki, Radiochim. Acta 40 (1986) 57.

\end{thebibliography}

\newpage

\newpage

\TableExplanation

\bigskip
\renewcommand{\arraystretch}{1.0}

\section*{Table 1.\label{tbl1te} Discovery of zinc, selenium, bromine, and neodymium isotopes }
\begin{tabular*}{0.95\textwidth}{@{}@{\extracolsep{\fill}}lp{5.5in}@{}}
\multicolumn{2}{p{0.95\textwidth}}{ }\\

Isotope & Zinc, selenium, bromine, or neodymium isotope \\
Author & First author of refereed publication \\
Journal & Journal of publication \\
Ref. & Reference \\
Method & Production method used in the discovery: \\
  & FE: fusion evaporation \\
  & LP: light-particle reactions (including neutrons) \\
  & MS: mass spectroscopy \\
  & NF: neutron induced fission \\
  & CPF: charged-particle induced fission \\
  & NC: neutron-capture reactions \\
  & PC: proton-capture reactions \\
  & PN: photo-nuclear reactions \\
  & SP: spallation reactions \\
  & PI: pion induced reactions \\
  & SF: spontaneous fission \\
  & PF: projectile fragmentation of fission \\
Laboratory & Laboratory where the experiment was performed\\
Country & Country of laboratory\\
Year & Year of discovery \\
\end{tabular*}
\label{tableI}

\datatables 



\setlength{\LTleft}{0pt}
\setlength{\LTright}{0pt}


\setlength{\tabcolsep}{0.5\tabcolsep}

\renewcommand{\arraystretch}{1.0}

\footnotesize 

\begin{longtable}{@{\extracolsep\fill}llllllll@{}}
\caption{Discovery of Zinc, Selenium, Bromine, and Neodymium Isotopes. See page\ \pageref{tbl1te} for Explanation of Tables}
Isotope & Author & Journal & Ref. & Method & Laboratory & Country & Year\\
\hline\\
\endfirsthead\\
\caption[]{(continued)}
Isotope & Author & Journal & Ref. & Method & Laboratory & Country & Year\\
\hline\\
\endhead
$^{54}$Zn & B. Blank & Phys. Rev. Lett. & \cite{2005Bla01}& PF & GANIL & France &2005 \\
$^{55}$Zn & J. Giovinazzo & Eur. Phys. J. A & \cite{2001Gio01}& PF & GANIL & France &2001 \\
$^{56}$Zn & J. Giovinazzo & Eur. Phys. J. A & \cite{2001Gio01}& PF & GANIL & France &2001 \\
$^{57}$Zn & D.J. Vieira & Phys. Lett. B & \cite{1976Vie01} & FE & Berkeley & USA &1976 \\
$^{58}$Zn & K.K. Seth & Phys. Lett. B & \cite{1986Set01}& PI & Los Alamos & USA &1986 \\
$^{59}$Zn & J. Honkanen & Nucl. Phys. A & \cite{1981Hon01}& LP & Jyvaskyla & Finland &1981 \\
$^{60}$Zn & L. Lindner & Physica & \cite{1955Lin01}& LP & Amsterdam & Netherlands &1955 \\
$^{61}$Zn & L. Lindner & Physica & \cite{1955Lin01}& LP & Amsterdam & Netherlands &1955 \\
$^{62}$Zn & D.R. Miller & Phys. Rev. & \cite{1948Mil01}& LP & Berkeley & USA &1948 \\
$^{63}$Zn & W. Bothe & Naturwiss. & \cite{1937Bot03}& NC & Heidelberg & Germany &1937 \\
$^{64}$Zn & A.J. Dempster & Phys. Rev. & \cite{1922Dem01}& MS & Chicago & USA &1922 \\
$^{65}$Zn & J.J. Livingood & Phys. Rev. & \cite{1939Liv03}& LP & Berkeley & USA &1939 \\
$^{66}$Zn & A.J. Dempster & Phys. Rev. & \cite{1922Dem01}& MS & Chicago & USA &1922 \\
$^{67}$Zn & F.W. Aston & Nature & \cite{1928Ast01}& MS & Cambridge & UK &1928 \\
$^{68}$Zn & A.J. Dempster & Phys. Rev. & \cite{1922Dem01}& MS & Chicago & USA &1922 \\
$^{69}$Zn & F.A. Heyn & Physica & \cite{1937Hey02}& NC & Eindhoven & Netherlands &1937 \\
$^{70}$Zn & A.J. Dempster & Phys. Rev. & \cite{1922Dem01}& MS & Chicago & USA &1922 \\
$^{71}$Zn & J.M. LeBlanc & Phys. Rev. & \cite{1955LeB01}& NC & Argonne & USA &1955 \\
$^{72}$Zn & J.M. Siegel & Nat. Nucl. Ener. Ser. & \cite{1951Sie01}& NF & Oak Ridge & USA &1951 \\
$^{73}$Zn & B.R. Erdal & Nucl. Phys. A & \cite{1972Erd01}& SP & CERN & Switzerland &1972 \\
$^{74}$Zn & B.R. Erdal & Nucl. Phys. A & \cite{1972Erd01}& SP & CERN & Switzerland &1972 \\
$^{75}$Zn & B. Grapengiesser & J. Inorg. Nucl. Chem. & \cite{1974Gra01}& NF & Studsvik & Sweden &1974 \\
$^{76}$Zn & B. Grapengiesser & J. Inorg. Nucl. Chem. & \cite{1974Gra01}& NF & Studsvik & Sweden &1974 \\
$^{77}$Zn & K. Aleklett & Nucl. Phys. A & \cite{1977Ale01}& NF & Studsvik & Sweden &1977 \\
$^{78}$Zn & K. Aleklett & Nucl. Phys. A & \cite{1977Ale01}& NF & Studsvik & Sweden &1977 \\
$^{79}$Zn & G. Rudstam & Nucl. Instrum. Meth. & \cite{1981Rud01}& NF & Studsvik & Sweden &1981 \\
$^{80}$Zn & G. Rudstam & Nucl. Instrum. Meth. & \cite{1981Rud01}& NF & Studsvik & Sweden &1981 \\
$^{81}$Zn & K.-L. Kratz & Z. Phys. A & \cite{1991Kra01} & SP & CERN & Switzerland &1991 \\
$^{82}$Zn & M. Bernas & Phys. Lett. B & \cite{1997Ber01}& PF & Darmstadt & Germany &1997 \\
$^{83}$Zn & M. Bernas & Phys. Lett. B & \cite{1997Ber01}& PF & Darmstadt & Germany &1997 \\
$^{84}$Zn & T. Ohnishi & J. Phys. Soc. Japan & \cite{2010Ohn01}& PF & RIKEN & Japan &2010 \\
$^{85}$Zn & T. Ohnishi & J. Phys. Soc. Japan & \cite{2010Ohn01}& PF & RIKEN & Japan &2010 \\
 &  &  &  &  &  & & \\
 &  &  &  &  &  & & \\
$^{64}$Se & A. Stolz & Phys. Lett. B & \cite{2005Sto01} & PF & Michigan State & USA &2005 \\
$^{65}$Se & J.C. Batchelder & Phys. Rev. C & \cite{1993Bat01} & FE & Berkeley & USA &1993 \\
$^{66}$Se & J.A. Winger & Phys. Rev. C & \cite{1993Win01} & PF & Michigan State & USA &1993 \\
$^{67}$Se & M.F. Mohar & Phys. Rev. Lett. & \cite{1991Moh01} & PF & Michigan State & USA &1991 \\
$^{68}$Se & C.J. Lister & Phys. Rev. C & \cite{1990Lis01} & FE & Daresbury& UK &1990 \\
$^{69}$Se & E. Nolte & Z. Phys. & \cite{1974Nol01} & FE & Munich & Germany &1974 \\
$^{70}$Se & H.H. Hopkins Jr. & Phys. Rev. & \cite{1950Hop01} & LP & Berkeley & USA &1950 \\
$^{71}$Se & J. Beydon & Compt. Rend. Acad. Sci. & \cite{1957Bey01} & FE & Saclay & France &1957 \\
$^{72}$Se & H.H. Hopkins Jr. & Phys. Rev. & \cite{1948Hop01}  & LP & Berkeley & USA &1948 \\
$^{73}$Se & W.S. Cowart & Phys. Rev. & \cite{1948Cow01} & LP & Ohio State & USA &1948 \\
$^{74}$Se & F.W. Aston & Nature & \cite{1922Ast03} & MS & Cambridge & UK &1922 \\
$^{75}$Se & H.N. Friedlander & Phys. Rev. & \cite{1947Fri01} & NC & Argonne & USA &1947 \\
$^{76}$Se & F.W. Aston & Nature & \cite{1922Ast03} & MS & Cambridge & UK &1922 \\
$^{77}$Se & F.W. Aston & Nature & \cite{1922Ast03} & MS & Cambridge & UK &1922 \\
$^{78}$Se & F.W. Aston & Nature & \cite{1922Ast03} & MS & Cambridge & UK &1922 \\
$^{79}$Se & A. Flammersfeld & Z. Naturforsch. & \cite{1950Fla01} & LP & Mainz & Germany &1950 \\
$^{80}$Se & F.W. Aston & Nature & \cite{1922Ast03} & MS & Cambridge & UK &1922 \\
$^{81}$Se & H. Waeffler & Helv. Phys. Acta & \cite{1948Waf01} & PN & Zurich & Switzerland &1948 \\
$^{82}$Se & F.W. Aston & Nature & \cite{1922Ast03} & MS & Cambridge & UK &1922 \\
$^{83}$Se & A.H. Snell & Phys. Rev. & \cite{1937Sne01} & LP & Berkeley & USA &1937 \\
$^{84}$Se & J.E. Sattizahn & J. Inorg. Nucl. Chem. & \cite{1960Sat01} & NF & Los Alamos & USA &1960 \\
$^{85}$Se & J.E. Sattizahn & J. Inorg. Nucl. Chem. & \cite{1960Sat01} & NF & Los Alamos & USA &1960 \\
$^{86}$Se & T. Tamai & Inorg. Nucl. Chem. Lett. & \cite{1973Tam01} & NF & Kyoto & Japan &1973 \\
$^{87}$Se & L. Tomlinson & J. Inorg. Nucl. Chem. & \cite{1968Tom01} & NF & Harwell & UK &1968 \\
$^{88}$Se & P. del Marmol & J. Inorg. Nucl. Chem. & \cite{1970Mar01} & NF & Mol & Belgium &1970 \\
$^{89}$Se & L. Tomlinson & J. Inorg. Nucl. Chem. & \cite{1971Tom01} & NF & Harwell & UK &1971 \\
$^{90}$Se & M. Bernas & Phys. Lett. B & \cite{1994Ber01} & PF & Darmstadt & Germany &1994 \\
$^{91}$Se & M. Asghar & Nucl. Phys. A & \cite{1975Asg01} & NF & Grenoble & France &1975 \\
$^{92}$Se & M. Bernas & Phys. Lett. B & \cite{1997Ber01} & PF & Darmstadt & Germany &1997 \\
$^{93}$Se & M. Bernas & Phys. Lett. B & \cite{1997Ber01} & PF & Darmstadt & Germany &1997 \\
$^{94}$Se & M. Bernas & Phys. Lett. B & \cite{1997Ber01} & PF & Darmstadt & Germany &1997 \\
$^{95}$Se & T. Ohnishi & J. Phys. Soc. Japan & \cite{2010Ohn01}& PF & RIKEN & Japan &2010 \\
 &  &  &  &  &  & & \\
 &  &  &  &  &  & & \\
$^{70}$Br & D.E. Alburger & Phys. Rev. C & \cite{1978Alb01} & FE & Brookhaven & USA &1978 \\
$^{71}$Br & B. Vosicki & Nucl. Instrum. Meth. & \cite{1981Vos01} & SP & CERN & Switzerland &1981 \\
$^{72}$Br & E. Nolte & Phys. Lett. B & \cite{1970Nol01} & FE & Heidelberg & Germany &1970 \\
$^{73}$Br & G. Murray & Nucl. Phys. A & \cite{1970Mur01} & FE & Manchester & UK &1970 \\
$^{74}$Br & J.M. Hollander & Phys. Rev. & \cite{1953Hol01} & FE & Berkeley & USA &1953 \\
$^{75}$Br & L.L. Woodward & Phys. Rev. & \cite{1948Woo02} & PC & Ohio State & USA &1948 \\
$^{76}$Br & S.C. Fultz & Phys. Rev. & \cite{1952Ful01} & LP & Ohio State & USA &1952 \\
$^{77}$Br & L.L. Woodward & Phys. Rev. & \cite{1948Woo02} & LP & Ohio State & USA &1948 \\
$^{78}$Br & W. Bothe & Naturwiss. & \cite{1937Bot01} & PN & Heidelberg & Germany &1937 \\
$^{79}$Br & F.W. Aston & Nature & \cite{1920Ast02} & MS & Cambridge & UK &1920 \\
$^{80}$Br & W. Bothe & Naturwiss. & \cite{1937Bot02} & PN & Heidelberg & Germany &1937 \\
$^{81}$Br & F.W. Aston & Nature & \cite{1920Ast02} & MS & Cambridge & UK &1920 \\
$^{82}$Br & A.H. Snell & Phys. Rev. & \cite{1937Sne01} & LP & Berkeley & USA &1937 \\
$^{83}$Br & A.H. Snell & Phys. Rev. & \cite{1937Sne01} & LP & Berkeley & USA &1937 \\
$^{84}$Br & H.J. Born & Naturwiss. & \cite{1943Bor01} & LP & Berlin & Germany &1943 \\
$^{85}$Br & H.J. Born & Naturwiss. & \cite{1943Bor01} & LP & Berlin & Germany &1943 \\
$^{86}$Br & A.F. Stehney & Phys. Rev. & \cite{1962Ste01} & LP & Argonne & USA &1962 \\
$^{87}$Br & H.J. Born & Naturwiss. &  \cite{1943Bor01} & LP & Berlin & Germany &1943 \\
$^{88}$Br & N. Sugarman & J. Chem. Phys. & \cite{1949Sug01} & NF & Argonne & USA &1948 \\
$^{89}$Br & G.J. Perlow & Phys. Rev. & \cite{1959Per01} & NF & Argonne & USA &1959 \\
$^{90}$Br & G.J. Perlow & Phys. Rev. & \cite{1959Per01} & NF & Argonne & USA &1959 \\
$^{91}$Br & K.L. Kratz & Nucl. Phys. A & \cite{1974Kra01} & NF & Mainz & Germany &1974 \\
$^{92}$Br & K.L. Kratz & Nucl. Phys. A & \cite{1974Kra01} & NF & Mainz & Germany &1974 \\
$^{93}$Br & B. Vosicki & Nucl. Instrum. Meth. & \cite{1981Vos01} & SP & CERN & Switzerland &1981 \\
$^{94}$Br & B. Vosicki & Nucl. Instrum. Meth. & \cite{1981Vos01} & SP & CERN & Switzerland &1981 \\
$^{95}$Br & M. Bernas & Phys. Lett. B & \cite{1997Ber01} & PF & Darmstadt & Germany &1997 \\
$^{96}$Br & M. Bernas & Phys. Lett. B & \cite{1997Ber01} & PF & Darmstadt & Germany &1997 \\
$^{97}$Br & M. Bernas & Phys. Lett. B & \cite{1997Ber01} & PF & Darmstadt & Germany &1997 \\
$^{98}$Br & T. Ohnishi & J. Phys. Soc. Japan & \cite{2010Ohn01}& PF & RIKEN & Japan &2010 \\
 &  &  &  &  &  & & \\
 &  &  &  &  &  & & \\
$^{125}$Nd & S.-W. Xu & Phys. Rev. C & \cite{1999Xu01} & FE & Lanzhou & China &1999 \\
$^{126}$Nd & & & & & & & \\
$^{127}$Nd & J.M. Nitschke & Z. Phys. A & \cite{1983Nit01} & FE & Berkeley & USA &1983 \\
$^{128}$Nd & C.J. Lister & Phys. Rev. Lett. & \cite{1985Lis01} & FE & Daresbury & UK &1985 \\
$^{129}$Nd & D.D. Bogdanov & Nucl. Phys. A & \cite{1977Bog01} & FE & Dubna & Russia &1977 \\
$^{130}$Nd & D.D. Bogdanov & Nucl. Phys. A & \cite{1977Bog01} & FE & Dubna & Russia &1977 \\
$^{131}$Nd & D.D. Bogdanov & Nucl. Phys. A & \cite{1977Bog01} & FE & Dubna & Russia &1977 \\
$^{132}$Nd & D.D. Bogdanov & Nucl. Phys. A & \cite{1977Bog01} & FE & Dubna & Russia &1977 \\
$^{133}$Nd & D.D. Bogdanov & Nucl. Phys. A & \cite{1977Bog01} & FE & Dubna & Russia &1977 \\
$^{134}$Nd & A.A. Abdurazakov & Izv. Akad. Nauk SSSR, Ser. Fiz & \cite{1970Abd01} & SP & Dubna & Russia &1970 \\
$^{135}$Nd & A.A. Abdurazakov & Izv. Akad. Nauk SSSR, Ser. Fiz & \cite{1970Abd01} & SP & Dubna & Russia &1970 \\
$^{136}$Nd & Zh. Zhelev & Izv. Akad. Nauk SSSR, Ser. Fiz & \cite{1968Zhe01} & SP & Dubna & Russia &1968 \\
$^{137}$Nd & Ch. Droste & Nucl. Phys. A & \cite{1970Dro01} & FE & Dubna & Russia &1970 \\
$^{138}$Nd & K. Ya. Gromov & Sov. Phys. JETP & \cite{1965Gro01} & SP & Dubna & Russia &1965 \\
$^{139}$Nd & B.J. Stover & Phys. Rev. & \cite{1951Sto01} & LP & Berkeley & USA &1951 \\
$^{140}$Nd & G. Wilkinson & Phys. Rev. & \cite{1949Wil01} & LP & Berkeley & USA &1949 \\
$^{141}$Nd & G. Wilkinson & Phys. Rev. & \cite{1949Wil01} & LP & Berkeley & USA &1949 \\
$^{142}$Nd & F.W. Aston & Nature & \cite{1924Ast02} & MS & Cambridge & UK &1924 \\
$^{143}$Nd & F.W. Aston & Nature & \cite{1933Ast01} & MS & Cambridge & UK &1933 \\
$^{144}$Nd & F.W. Aston & Nature & \cite{1924Ast02} & MS & Cambridge & UK &1924 \\
$^{145}$Nd & F.W. Aston & Nature & \cite{1933Ast01} & MS & Cambridge & UK &1933 \\
$^{146}$Nd & F.W. Aston & Nature & \cite{1924Ast02} & MS & Cambridge & UK &1924 \\
$^{147}$Nd & J.A. Marinsky & J. Am. Chem. Soc. & \cite{1947Mar01} & NF & Oak Ridge & USA &1947 \\
$^{148}$Nd & A.J. Dempster & Phys. Rev. & \cite{1937Dem02} & MS & Chicago & USA &1937 \\
$^{149}$Nd & M.L. Pool & Phys. Rev. & \cite{1938Poo02} & LP & Michigan & USA &1938 \\
$^{150}$Nd & A.J. Dempster & Phys. Rev. & \cite{1937Dem02} & MS & Chicago & USA &1937 \\
$^{151}$Nd & M.L. Pool & Phys. Rev. & \cite{1938Poo02} & LP & Michigan & USA &1938 \\
$^{152}$Nd & A. Wakat & Radiochem. Radioanal. Lett. & \cite{1969Wak01} & NF & Michigan & USA &1969 \\
$^{153}$Nd & R.C. Greenwood & Phys. Rev. C & \cite{1987Gre01} & SF & Idaho Falls & USA &1987 \\
$^{154}$Nd & J.B. Wilhelmy & Phys. Rev. Lett. & \cite{1970Wil01} & SF & Berkeley & USA &1970 \\
$^{155}$Nd & K. Okano & Radiochim. Acta & \cite{1986Oka01} & NF & Kyoto & Japan &1986 \\
$^{156}$Nd & R.C. Greenwood & Phys. Rev. C & \cite{1987Gre01} & SF & Idaho Falls & USA &1987 \\
 \\
\end{longtable}

\end{document}